%% file: metrics.tex
\documentclass[paper, 12pt, letterpaper, epsf]{JHEP} 
\input{epsf.tex}
\usepackage{graphics} 
\usepackage{epsfig}

\def\Bbb{\bf} 

\def\C{{\Bbb C}} 
\def\R{{\Bbb R}}
\def\Z{{\Bbb Z}} 
\def\H{{\Bbb H}} 

\def\P{{\Bbb P}}
\def\i{{\Bbb i}}\def\j{{\Bbb j}}\def\k{{\Bbb k}}

    \def\det{{\rm det}}

\def\id{\protect{{1 \kern-.28em {\rm l}}}}

\newcommand{\be}{\begin{equation}} \newcommand{\ee}{\end{equation}}
\newcommand{\bea}{\begin{eqnarray}} \newcommand{\eea}{\end{eqnarray}}
\newcommand{\beann}{\begin{eqnarray*}} \newcommand{\eeann}{\end{eqnarray*}}
\newcommand{\bfig}{\begin{figure}} \newcommand{\efig}{\end{figure}}
\newcommand{\nn}{\nonumber}
\newcommand{\ba}{\begin{array}}\newcommand{\ea}{\end{array}}

\newtheorem{Proposition}{Proposition}[section]

\newtheorem{Theorem}{Theorem}[section]
\newtheorem{Lemma}{Lemma}[section]
\newtheorem{Corrolary}{Corrolary}[section]

\newcommand{\bp}{\begin{Proposition}} \newcommand{\ep}{\end{Proposition}} 
\newcommand{\bt}{\begin{Theorem}} \newcommand{\et}{\end{Theorem}} 
\newcommand{\bl}{\begin{Lemma}} \newcommand{\el}{\end{Lemma}} 
\newcommand{\bc}{\begin{Corrolary}} \newcommand{\ec}{\end{Corrolary}} 

\title{M-theory on toric 
$G_2$ cones and its type II reduction.}

\author{L. Anguelova, C. I. Lazaroiu
\\C.~N.~Yang Institute for Theoretical Physics\\
SUNY at Stony BrookNY11794-3840,
U.S.A.\\anguelov, calin @insti.physics.sunysb.edu}

\abstract{We analyze a class of conical $G_2$ metrics admitting 
two commuting isometries, together with a certain one-parameter family of  
$G_2$ deformations which preserves these symmetries. 
Upon using recent results of Calderbank and Pedersen, 
we extract the IIA reduction of M-theory 
on such backgrounds, as well as its type IIB dual.
The associated type II solutions are expected to contain 6-branes and 
5-branes respectively.
By studying the general asympotics of the IIA and IIB 
solutions around the relevant loci, we confirm the interpretation 
of such backgrounds in terms of localized and delocalized branes. 
In particular, we find explicit, general expressions for the string coupling 
and R-R/NS-NS fields in the vicinity of these objects. Our solutions contain 
and generalize the field configurations relevant for 
certain models considered in recent work of Acharya and Witten. }

\preprint{YITP-SB-02-23}

\begin{document}

\tableofcontents

\pagebreak 

\vskip .6in

\section{Introduction} 

M-theory on backgrounds of $G_2$ holonomy has recently 
attracted renewed attention \cite{Brandhuber, Cvetic, CGLP, 
Gukov, GT, GYZ, KB, Witten_anom, Witten_Acharya, AW}. An interesting class of such models
arises from certain non-compact $G_2$ spaces obtained by 
a construction due to \cite{BS} and \cite{GP}. These result as 
cones ${\cal C}(Y)$ built over the twistor space $Y$ of a compact,  
Einstein self-dual space $M$ of positive scalar curvature. In this 
construction, $Y$ is endowed with a `twisted' metric, which differs from 
its canonical Kahler-Einstein metric by certain global and fiberwise 
rescalings. If one restricts to the case of smooth $M$, then there are 
precisely two choices\footnote{These can be recognized as two of the 
Wolf-Alekseevskii spaces \cite{Wolf, Aleks1,Aleks2}, namely 
$S^4=\H\P^1=Sp(2)/[Sp(1)\times Sp(1)]$ and $M=\C\P^2=SU(3)/U(2)=Gr_2(\C^3)$,
the Grassmannian of two planes in $\C^3$. Also note that $\C\P^3=
Sp(2)/[U(1)\times Sp(1)]$.}
(namely $M=S^4$ and $M=\C\P^2$, for which 
$Y$ is respectively $\C\P^3$ and $SU(3)/U(1)^2$), both of which were 
analyzed in \cite{AW}. Many more models can be obtained by allowing $M$ to be 
an orbifold. As pointed out in \cite{Witten_anom} and \cite{Witten_Acharya},
such backgrounds 
are especially interesting from a phenomenological perspective, 
since their low energy effective description gives four-dimensional, 
nonabelian gauge theories containing chiral matter. A particular class of 
such models (namely those based on $M=W\C\P^2_{p,q,r}$, endowed with the 
ESD metrics obtained implicitly in \cite{GL}) was analyzed 
in \cite{Witten_Acharya}. In this case, the authors 
show that 
the type IIA reduction leads to configurations of three 
intersecting D6-branes $B_1, B_2,B_3$, 
each carrying a nonabelian $SU(m_j)$ gauge theory, with $m_j$ determined 
by the integral weights $p,q$ and $r$. The phenomenological relevance 
of models based on intersecting branes 
is underscored by the work of \cite{CSU1,CSU2}.

The models considered in \cite{Witten_Acharya} share one 
distinguishing feature. Namely, the ESD metric on $M$ admits a two-torus of 
isometries, which lift to isometries of the $G_2$ metric on ${\cal C}(Y)$
through the construction of \cite{GP} and \cite{BS}. Therefore, 
one obtains a natural 
generalization by allowing $M$ to be an ESD orbifold 
(compact and of positive scalar curvature) admitting a two-torus of isometries.
M-theory on such cones was considered in \cite{toric}. There, 
we presented an algorithm for identifying the type and location of 
singularities  of the $G_2$ cone ${\cal C}(Y)$.
The work of \cite{toric} (which we briefly review in Section 2), relies on 
using a well-known correspondence between ESD  spaces and hyperkahler 
cones \cite{Swann, BGMR,LeBrun,LeBrun_finite,BG}. 
In the case under consideration, this
construction produces a {\em toric hyperkahler} \cite{BD} cone, 
which can be studied with methods reminiscent of toric geometry
\cite{Oda,Fulton,Danilov,Audin, Cox,Cox_review}.  
In \cite{toric}, we used this observation in order 
to give general rules for finding the singularities of ${\cal C}(M)$, 
and thus the low energy gauge group of M-theory on such backgrounds. 
Since the $G_2$ cones considered in \cite{toric} 
admit  an isometric $T^2$ action, 
it is clear that such systems admit T-dual type IIA and type 
IIB interpretations. By using somewhat abstract 
arguments, we found that the type IIA reduction of these models 
will generally contain strongly coupled 6-branes, 
and that the type IIB description is typically given in terms of 
(strongly coupled) delocalized 5-branes. 

While these results suffice for a qualitative understanding of the physics, 
one is left wondering about the  11d supergravity and IIA/IIB 
solutions associated with our models (of which the examples of 
\cite{Witten_Acharya} are a particular case). It turns out that the 
solutions of interest can be determined explicitly.
This follows from a recent result of Calderbank and Pedersen \cite{CP},
who used an elegant chain of arguments in order to write down the most 
general ESD metric admitting two independent and commuting Killing fields.\footnote{Particular cases of such metrics were considered in \cite{CIV1} and \cite{CIV2}.} 
For the particular case of positive scalar curvature, they also give the 
explicit expression of this solution in terms of the
toric hyperkahler data which play a central role in \cite{toric}.  

In the present paper, we combine results of \cite{CP} and \cite{BS, GP} 
in order to extract the explicit 11d $G_2$ holonomy metric associated with 
this class of models. Upon reducing through one of the 
$T^2$ isometries, we will be able to 
obtain the associated type IIA backgrounds, as well as their T-dual counterpart 
in IIB. This gives a discrete infinity of 
IIA and IIB  solutions for certain configurations of 
localized and delocalized branes which preserve $N=1$ spacetime 
supersymmetry. 
Moreover, the asymptotics of the relevant fields around the brane
locations provides an independent confirmation of the interpretation 
proposed in \cite{toric}.

The present paper is organized as follows. 
In Section 2, we give a brief review 
of the models of interest and of some basic results of \cite{toric}. 
Section 3 summarizes the result of \cite{CP}, and explains how it relates
to the picture of our previous paper. 
In Section 4, we perform the IIA reduction 
of the resulting $G_2$ metrics and extract the physical interpretation of 
the solution by analyzing the asymptotics of various fields around 
a certain locus. We also explain how this analysis relates to that of 
\cite{toric}. Section 5 considers the T-dual, type IIB solution 
and its physical interpretation. In Section 6, we give a general 
expression for the calibration 3-form of our models. 
Section 7 presents our conclusions.

\section{$G_2$ cones from ESD orbifolds.}

It is well-known \cite{Sal,Bergery} that to every $4d$-dimensional
quaternion-Kahler space $M$ one can associate a $4d+2$-dimensional
twistor space $Y$ and (modulo the possibility of a double cover)
\cite{Swann, BG,LeBrun,LeBrun_finite,MR} 
a $4(d+1)$-dimensional hyperkahler cone\footnote{By
definition a hyperkahler cone is a hyperkahler
space which can be written as the metric cone over a compact
Riemannian space.} $X$. This allows one to present the 
twistor space of $M$ as a Kahler 
quotient of the hyperkahler cone at some positive moment map level.
For $d=1$, this construction associates an eight-dimensional hyperkahler 
cone with every compact, ESD space of positive scalar curvature, and 
reduces the study of its twistor space to that of a certain Kahler quotient 
of this cone.

On the other hand, the papers \cite{BS} and \cite{GP} 
construct a $G_2$ cone from each ESD space of positive curvature\footnote{
These papers consider the case when $M$ is smooth, but their results
admit an obvious generalization to the orbifold case.}. 
For an ESD space, the twistor space $Y$ can be described as
the bundle of unit anti-self-dual two-forms on $M$:
\be
Y=B_1(\Lambda^{2,-}(T^*M))~~.
\ee
This carries the Kahler-Einstein metric: 
\be
\label{rho}
d\rho^2=|d\sigma|^2+|d_A{\vec u}|^2~~, 
\ee
where $d\sigma^2$ is the self-dual Einstein metric on $M$, 
${\vec u}=(u^1,u^2,u^3)$ are coordinates with respect to a local
frame of sections of $\Lambda^{2,-}(T^*M)$ (subject to the constraint 
$|\vec{u}|^2 = 1$) and the connection $A$ on this bundle is 
induced by the Levi-Civita connection of $M$.

Following \cite{GP,BS}, we consider the `modified' metric on $Y$:
\be
\label{rhoprime}
d\rho'^2=\frac{1}{2}\left[d\sigma^2+\frac{1}{2}|d_A{\vec u}|^2\right]~~, 
\ee
and construct a $G_2$ cone ${\cal C}(Y)$ 
as the metric cone over $Y$, taken with respect 
to this metric:
\be
\label{G2metric}
ds^2=dr^2+r^2d\rho'^2=dr^2+\frac{r^2}{2}(d\sigma^2+
\frac{1}{2}|d_A{\vec u}|^2)~~.
\ee
This metric admits a one-parameter family of $G_2$ deformations:
\be
\label{G2deformed}
ds^2=\frac{1}{1-(r_0/r)^4}dr^2+\frac{r^2}{2}
(d\sigma^2+\frac{1}{2}(1-(r_0/r)^4)|d_A{\vec u}|^2)~~,~~r_0\geq 0~~,
\ee
with the conical case obtained for $r_0 = 0$.

The papers \cite{AW} and \cite{Witten_Acharya} studied 
$M$-theory on such $G_2$ backgrounds for $M$ taken to be 
$S^4$ or $\C \P^2$ (these are the only smooth choices) 
and for $M$ given by a weighted projective space, endowed with the orbifold 
ESD  metrics constructed indirectly in \cite{GL}.

It is now well-established \cite{GL, AG, CP, BG, BGMR} that there exist
infinitely many inequivalent Einstein self-dual orbifolds of positive scalar 
curvature, of which the examples mentioned above form only a small subclass.
While the most general model of this type is still beyond reach, there exists 
a natural class of examples which includes the cases $M=W\C\P^2_{p,q,r}$ 
while allowing for a wide generalization. This is the class 
of `toric' ESD orbifolds, i.e. ESD spaces which possess a two-torus of 
isometries. As explained in \cite{toric}, the hyperkahler cone of 
such a space is `toric hyperkahler' in the sense of \cite{BD}\footnote{We warn the reader that the terminology `toric hyperkahler' 
was used in a different sense in 
the work of \cite{GGPT,G,GR}. In this paper, we exclusively 
use this terminology in the sense of \cite{BD}.}, an 
observation which allows 
for a systematic analysis of the singularities of its twistor space and  
associated $G_2$ cone. Moreover, all ESD metrics of this type 
are known explicitly due to recent work of Calderbank and Pedersen \cite{CP}. 

For such models, the isometries of the
four-dimensional base $M$ lift to isometries of the twistor space
metrics (\ref{rho}) and (\ref{rhoprime}). From (\ref{G2deformed}), we also 
find that the latter lift to isometries of the
$G_2$ cone and its one parameter deformations. Following \cite{toric}, 
$G_2$ isometries of this type will be called {\em special}.
As explained in \cite{toric}, there exists a correspondence 
between `compact' special isometries and vectors of the $\Z^2$ lattice 
spanned by the toric hyperkahler generators $\nu_j$.  
M-theory on our $G_2$
backgrounds can be reduced to type IIA upon quotienting through a fixed 
special isometry. This leads to a type IIA background which admits a 
`compact' Killing vector field (which generates a 
$U(1)$ group of isometries), inherited 
from the two-torus of special isometries upon reduction. This allows one 
to extract a T-dual type IIB description. It follows that all models 
of this type admit both IIA and IIB interpretations, namely a pair of such 
for each choice of special isometry. Before studying these reductions, we
review some results of \cite{toric} which will help us extract their 
interpretation.

\subsection{Eight-dimensional toric hyperkahler cones}

\setcounter{equation}{0}
 
As mentioned above, the hyperkahler cones associated with our models are 
{\em toric hyperkahler} in the sense of \cite{BD}. This means 
that they are given by torus 
hyperkahler quotients \cite{HKLR} of the form $X=\H^n///_0U(1)^{n-2}$ 
of some quaternion affine space $\H^n$ at zero levels of the hyperkahler 
moment map. The $U(1)^{n-2}$ action has the following form on  
affine quaternion coordinates $u_1\dots u_n\in \H$:
\be
\label{action}
u_k \rightarrow 
\prod_{\alpha=1}^{n-2}{\lambda_\alpha^{q_k^{(\alpha)}}}u_k~~, 
\ee
where $\lambda_{\alpha}$ are complex numbers of unit modulus and 
$q_k^{(\alpha)}$ are some integers.  
The latter form an $(n-2)\times n$ matrix $Q_{\alpha k}=q_k^{(\alpha)}$.
To obtain an effective $U(1)^{n-2}$ action on $\H^n$, one requires that this 
matrix has unit invariant factors, which amounts to asking that the 
map of lattices $q^*:\Z^{n-2}\rightarrow \Z^n$ associated with the transpose
matrix $Q^t$ has torsion-free cokernel. In this case, one obtains a 
short exact sequence:
\be
\label{es}
0\longrightarrow
\Z^{n-2}\stackrel{q^*}{\longrightarrow}\Z^n\stackrel{g}{\longrightarrow}
\Z^2\longrightarrow 0~~,
\ee
where the map $g$ is represented by a $2\times n$ integral matrix $G$, 
whose columns $\nu_j=g(e_j)\in \Z^2$ are the so-called 
{\em toric hyperkahler generators} (here $e_j$ is the 
canonical basis of $\Z^n$). The paper \cite{BD} focuses on the case 
of primitive vectors\footnote{Recall that a vector is called primitive if its components are relatively prime integers.} $\nu_j$ and, in fact, considers mostly the case of 
non-vanishing moment map levels (i.e. studies spaces of the type 
$X=\H^n///_{\xi}U(1)^m$ for certain nonzero levels $\xi$). 
The relevant results 
of \cite{BD} were extended to our situation in the paper \cite{toric}. 

Upon introducing complex coordinates $w_k^{(\pm)}$ through:
\be
u_k = w_k^{(+)} + \j w_k^{(-)} ~~,
\ee
(where $\i,\j,\k$ are the imaginary quaternion units, $\i$ being identified 
with $\sqrt{-1}$), the torus action (\ref{action}) becomes:
\be
\label{complex_action}
w_k^{(+)}\rightarrow \prod_{\alpha=1}^{n-2} 
\lambda_\alpha ^{q^{(\alpha)}_k}w_k^{(+)}~~,~~
w_k^{(-)}\rightarrow \prod_{\alpha=1}^{n-2} 
\lambda_\alpha ^{-q^{(\alpha)}_k}w_k^{(-)}~~,
\ee
while the $U(1)$ action leading to the twistor space $Y$ has the form:
\be
\label{U1proj}
w_k^{(\pm)}\rightarrow \lambda w_k^{(\pm)}~~.
\ee

As in \cite{BD}, the hyperkahler 
cone admits a tri-Hamiltonian action by the two-torus
$U(1)^2=U(1)^n/U(1)^{n-2}$, which allows us to write $X$ as a $T^2$ fibration 
over $\R^6=(\R^3)^2$. The fibration map ${\vec \pi}=
({\vec \pi}^{(1)}, {\vec \pi}^{(2)}):X\rightarrow \R^3\times \R^3$ is the 
hyperkahler moment map for this $U(1)^2$ action. The fibers of ${\vec \pi}$ 
can be described explicitly as follows. If one introduces appropriate 
vector coordinates 
${\vec x}=(x_1,x_2,x_3)$ and ${\vec y}=(y_1,y_2,y_3)$ for the two $\R^3$ 
factors, then the fiber above 
$({\vec x}, {\vec y})$ consists of solutions 
$u=(\{w_k^{(+)}\}, \{w_k^{(-)}\})$ to the system:
\be
\label{ab}
\frac{1}{2}(|w^{(+)}_k|^2-|w^{(-)}_k|^2)=\nu_k\cdot a~~,~~
w^{(+)}_kw^{(-)}_k=\nu_k\cdot b~~,
~{\rm~for~all~}k=1\dots n~~, \label{X}
\ee
where $a=(x_1, y_1)\in \R^2$ and 
$b=(x_3+ix_2, y_3+iy_2)\in \C^2$. In (\ref{ab}), the symbol 
$\cdot$ stands for the scalar product. 

\subsection{The distinguished locus}
\setcounter{equation}{0}

It is clear that the singularities of our $G_2$ metrics are immediately 
known given the singularities of $Y$. The latter were studied in \cite{toric} 
by using the Kahler quotient description $Y=X//_{\zeta}U(1)$, where $\zeta>0$ 
is some level which specifies the overall scale of $Y$ 
and $U(1)$ is the action (\ref{U1proj}). In that paper, 
we showed that all singularities of the twistor space lie on the so-called 
{\em distinguished locus} $Y_D\subset Y$, which is a certain union of 
holomorphically embedded two-spheres in $Y$. The locus $Y_D$ can 
be defined abstractly as follows. The $T^2$ fibration $X\rightarrow M$ 
descends to a $T^2$ fibration $Y\rightarrow M$ after performing the 
$U(1)$ Kahler quotient. Then $Y_D$ is the locus in $Y$ where the 
fibers of this map collapse to a circle or to a point. 

\begin{figure}[hbtp]
\begin{center}
\scalebox{0.4}{\input{dist.pstex_t}}
\end{center}
\caption{\label{dist} Fibration of the distinguished locus over the polygon 
$\Delta$.}
\end{figure}
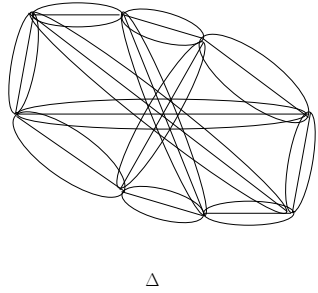

The distinguished locus is 
most easily described in terms of the {\em characteristic polygon}
of $Y$, which is the convex two-dimensional polygon $\Delta$ defined through:
\be
\Delta=\{a\in \R^2|\sum_{k=1}^n{|\nu_k\cdot a|}=\zeta\}~~.
\ee
This polygon has $2n$ vertices and is symmetric with respect to 
the reflection $\iota :a\rightarrow -a$ through the origin of $\R^2$
(figure \ref{DDM}); its 
{\em principal diagonals} (i.e. those diagonals passing through the origin)
have the form $D_k:=\{a\in \R^2|\nu_k\cdot a=0\}$, where $k$ runs from $1$ 
to $n$.  
In \cite{toric}, we show that the map $u\rightarrow a$ given in 
(\ref{ab}) presents $Y_D$ as an $S^1$ fibration over the one-dimensional 
space obtained as the union of all edges and principal diagonals of $\Delta$
(see figure \ref{dist}).
The $S^1$ fibers collapse to points above the vertices of the characteristic 
polygon. This fibration leads to a natural decomposition of the distinguished
locus into a {\em horizontal component} $Y_H$ (the part of $Y_D$ which lies 
above the edges of $\Delta$) and a {\em vertical component} $Y_V$ 
(the part of $Y_D$ lying above  the principal diagonals). These further 
decompose as $Y_H=\cup_{e\in E}{Y_e}$ and $Y_V=\cup_{k=1}^n{Y_k}$, 
where $E$ is the set of edges of $\Delta$ and $Y_e$, $Y_j$ are the two-spheres 
lying above the edge $e$ and above 
the principal diagonal $D_j$. To describe these
two-spheres explicitly, one introduces sign vectors 
$\epsilon(e)=(\epsilon_1(e)\dots \epsilon_n(e))$ associated with the edges 
$e$, whose components are defined by:
\be
\epsilon_j(e)=sign(\nu_j\cdot p_e)~~,
\ee
where $p_e$ is the vector associated with the middle of $e$ (figure 
\ref{DDM}). 
Then $Y_e$ is the locus in $Y$ defined by simultaneous vanishing of the complex
coordinates $w^{(-\epsilon_1(e))}_1\dots w^{(-\epsilon_n(e))}_n$, while 
$Y_j$ is defined by vanishing of the quaternion coordinate $u_j$, i.e. 
by simultaneous vanishing of the complex coordinates $w_j^{(+)}$ and 
$w_j^{(-)}$. It is also shown in \cite{toric} that the spheres $Y_j$ are 
fibers of the $S^2$ fibration $Y\rightarrow M$ (hence the name 
`vertical component' for $Y_V$ ) , while $Y_e$ are horizontal with respect to 
this fibration (i.e. are lifts of spheres lying in the ESD base  $M$).
The twistor space admits an involution which commutes with the projection 
$Y\rightarrow M$, and thus acts along its $S^2$ fibers. As discussed in 
\cite{toric}, the restriction of this `antipodal map' to the horizontal 
locus covers the involution $\iota :a\rightarrow -a$
of $\Delta$ through the projection map $Y_H\rightarrow \Delta$. 
This implies that the singularity types of $Y$ along 
$Y_e$ and $Y_{-e}$ coincide, where $e$ and $-e$ are opposite edges of 
$\Delta$. It also shows that the components $Y_e, Y_{-e}$ of the 
horizontal locus project to the same locus in the ESD base $M$. 
The collection of such projections is associated with a 
polygon $\Delta_M$ on $n$ vertices, obtained by quotienting $\Delta$ through 
the sign inversion $\iota$ (figure \ref{DDM}):
\be
\Delta_M=\Delta/\iota~~.
\ee
This polygon (which also appeared in the work of \cite{BGMR} and 
\cite{CP}, though from a 
different perspective) will play an important role in what follows.

\begin{figure}[hbtp]
\begin{center}
\scalebox{0.4}{\input{DDM.pstex_t}}
\end{center}
\caption{\label{DDM} Examples of the 
polygons $\Delta$ and $\Delta_M$ for $n=4$. For the edge $e$ of $\Delta$ 
drawn as a bold line, we show the vector $p_e$ used in the definition of
the signs $\epsilon_j(e)$. Note that the principal diagonals 
$D_j=\{a|a\cdot \nu_j=0\}$ need not lie in trigonometric order.}
\end{figure}
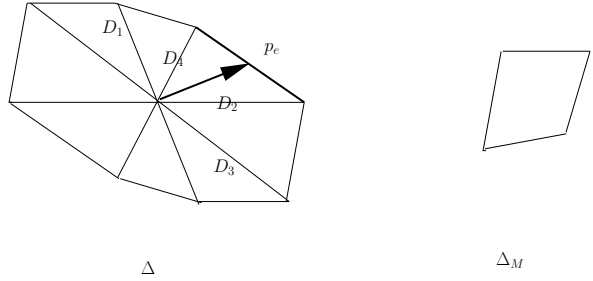

Assuming that the $\Z_2$ subgroup of $U(1)$ acts nontrivially on 
$X$ (see (\ref{U1proj})), it 
was shown in \cite{toric} that $Y$ has a $\Z_{m_e}$ singularity along 
$Y_e$, where $m_e$ is the greatest common divisor of the two components of 
the integral vector:
\be
\label{nu_e}
\nu_e:=\sum_{k=1}^n{\epsilon_k(e)\nu_k}~~
\ee
(the generic point of $Y_e$ is smooth if $m_e$ equals one). 
The singularity type along $Y_j$ is given by $\Z_{m_j}$ or $\Z_{2m_j}$, 
where $m_j$ is the greatest common divisor of the components of the 
toric hyperkahler generator $\nu_j$. Which of the cases 
$\Z_{m_j}$ or $\Z_{2m_j}$ occurs along each $Y_j$ 
can be determined by a simple criterion.
These singularity types can be enhanced at the intersection points between 
$Y_e$ and $Y_j$, which lie above the vertices of $\Delta$. 
If the $\Z_2$ subgroup of $U(1)$ acts trivially on $X$, then the singularity 
type along $Y_j$ is always given by $\Z_{m_j}$, while the 
singularity type along $Y_e$ is `half' of that determined above 
(namely in this case $m_e$ is even, and the singularity type is 
$\Z_{m_e/2}$). 

\section{Toric ESD metrics and $G_2$ metrics}

Consider an ESD space $M$ whose associated hyperkahler cone is defined by 
the exact sequence (\ref{es}).
To describe the result of \cite{CP}, we 
introduce coordinates $\phi, \psi$ on the $T^2$ fiber of $X\rightarrow
\R^6$.  Recall that a point in $\R^6=(\R^3)^2$ has the form
$({\vec x}, {\vec y})$, where ${\vec x}=(x_1,x_2,x_3)$ and ${\vec
y}=(y_1,y_2,y_3)$ are real 3-vectors.
We denote by ${\cal H}^2$ the hyperbolic plane (upper half plane), 
with coordinates
$\eta\in \R$ and $\rho\geq 0$, and by ${\overline {\cal H}}^2$ its one-point 
compactification. The latter is the disk obtained by adding the point 
$|\eta|=\infty$. Following \cite{CP}, we define a
surjective function $\Theta:\R^3\times \R^3\rightarrow {\overline {\cal
H}}^2$ by: 
\be
\label{Theta}
\rho=\frac{|{\vec x}\times {\vec y}|^2}{|{\vec x}|^2}~~,~~
\eta=\frac{{\vec x}\cdot {\vec y}}{|{\vec x}|^2}~~.  
\ee 
It is shown in \cite{CP} that the maps in figure \ref{CO} form a commutative
diagram, whose upper row is the projection $X\rightarrow M$ induced by
performing a certain $\H^*$ quotient\footnote{The group $\H^* = \H
\backslash \{0\}$ has a natural action on the toric hyperkahler cone 
$X$, induced by the action on $\H^n$ which takes $u_k$ into $u_k t^{-1}$ 
for all $k=1\dots n$ (here $t$ is an element of $\H^*$).
As reviewed in \cite{toric}, 
the $\H^*$ quotient of $X$ recovers the quaternion
space $M$. This reduction is a particular case of the 
so-called 
`superconformal quotient' of \cite{conf_quotient,martin,martin_review}.}. 
In fact, the surjection $\pi_M$ on the
right presents $M$ as a fibration over the upper half plane, whose generic 
fiber is a two-torus. The coordinates $\phi,\psi$ descend to 
coordinates on the $T^2$ fiber of $M$. Hence
$(\eta,\rho,\phi, \psi)$ give a coordinate system on $M$, adapted to
its $T^2$ fibration over ${\overline {\cal H}}^2$.

\begin{figure}[hbtp]
\begin{center}
\scalebox{0.5}{\input{CO.pstex_t}}
\end{center}
\caption{Performing the $\H^*$ quotient.\label{CO}}
\end{figure}
 
Let us define a function on ${\cal H}^2$ by:
\be
\label{F}
F=\sum_{k=1}^n \frac{\sqrt{(\nu_k^2)^2\rho^2
+(\nu_k^2\eta+\nu_k^1)^2}}{\sqrt\rho}~~.  
\ee 
It is then shown in
\cite{CP} that the Einstein self-dual metric on $M$ has the form:
{\footnotesize \bea \!\!\!\!\!\!\!\!\!\!  d\sigma^2=\frac{ F^2 -
4\rho^2(F_\rho^2 + F_\eta^2) }{4 F^2}\; \frac{d\rho^2 +
d\eta^2}{\rho^2} +\frac{ \left[ (F - 2 \rho F_\rho) \alpha - 2 \rho
F_\eta \beta \right]^2 + \left[ -2\rho F_\eta \alpha + (F + 2 \rho
F_\rho )\beta \right]^2 }{ F^2\left[F^2-4\rho^2(F_\rho^2 +
F_\eta^2)\right]}, \label{Mmetric} \eea} where
$\alpha=\sqrt\rho\,d\phi$, $\beta=(d\psi+\eta\, d\phi)/\sqrt\rho$ and
$F_{\rho}= \partial F/\partial \rho$.

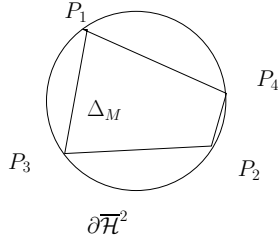
\begin{figure}[hbtp]
\begin{center}
\scalebox{0.5}{\input{sph.pstex_t}}
\end{center}
\caption{\label{sph} The polygon $\Delta_M$ is homeomorphic with the boundary 
of ${\overline {\cal H}}^2$, but not diffeomorphic with it. The figure 
shows the case $n=4$.}
\end{figure}

Using topological methods, the authors of \cite{BGMR} prove that the
$T^2$ fibers of $M$ collapse to zero size above the boundary $\rho=0$
of ${\overline {\cal H}}^2$. 
From (\ref{Theta}), this boundary corresponds to the
region of $\R^6$ where ${\vec x}$ and ${\vec y}$ become
linearly dependent. At a generic point on $\partial {\overline {\cal
H}}^2$, one of the cycles of the $T^2$ fiber collapses to a point. The
entire two torus collapses to a point at special positions on the
boundary \cite{BGMR}. As we shall see in a moment, the boundary $\partial {\overline {\cal H}}^2$ can be identified topologically with the polygon $\Delta_M$, in such a way that the vertices of $\Delta_M$ correspond to these special points (figure \ref{sph}).

As recalled above, the $T^2$ fiber of $Y\rightarrow M$
collapses along the distinguished locus $Y_D$. 
The same must
obviously be true for the projection of this locus to the self-dual manifold
$M$. Hence the distinguished locus projects to a locus $M_D$ in $M$, 
which sits above the boundary of ${\overline {\cal
H}}^2$. To understand this
projection explicitly, 
consider first a vertical component $Y_k$, which corresponds to 
$\nu_k\cdot a=\nu_k\cdot b=0$ i.e.
$\nu_k^1{\vec x}+\nu_k^2{\vec y}=0$. On this subset of  $\R^6$, 
one has $\rho=0$ and $\eta=- \frac{\nu_k^1}{\nu_k^2}$, which
gives points $P_k$ ($k=1\dots n$) lying on $\partial {\overline {\cal H}}^2$.
A horizontal component $Y_e$ corresponds to
$b=0\Leftrightarrow x_2=x_3=y_2=y_3=0$ and a certain choice $\epsilon_k(e)$ 
for the signs of the scalar products 
$\nu_k\cdot a=\nu_k^1x_1+\nu_k^2 y_1$. This gives $\rho=0$
and a range for the slope $\eta=\frac{y_1}{x_1}$. Therefore, one 
obtains  a segment on $\partial {\overline {\cal H}}^2$, 
connecting two of the points
$P_k$.  It is also clear that the locus $Y_{-e}$ associated with the
opposite edge of $\Delta$ gives the same segment on this
boundary. 
Putting everything together, we find that $\partial {\overline {\cal H}}^2$ 
can be identified with the polygon $\Delta_M$ of the 
previous subsection, such that the points $P_k$ correspond to its vertices (see figure \ref{sph}).
The $T^2$ fiber of $M$ collapses to a circle above the interior of 
the edges of $\Delta_M$ and to a point above its vertices. This
picture agrees with that of \cite{BGMR}. In this paper, we shall freely use the identification $\Delta_M \equiv \partial {\overline {\cal H}}^2$, therefore identifying the edges of $\Delta_M$ with the circular segments $(P_k, P_{k+1})$.

\paragraph{Observation 1} Note 
that the circular segment $(P_k,P_{k+1})$ on the boundary of 
${\overline {\cal H}}^2$ is characterized by the condition $\rho=0$ 
and a choice for the signs 
$\epsilon_j(\eta):=sign(\nu_j^2\eta+\nu_j^1)$.
Indeed, the quantity $\nu_k^2\eta+\nu_k^1$ changes sign precisely at the 
point $P_k$ (given by $\eta=\eta_k:=-\frac{\nu_k^1}{\nu_k^2}$). Therefore, the 
signs $\epsilon_j(\eta)$ must be constant on each of the (open) 
circular segments $(P_k,P_{k+1})$, and the sign vector 
$\epsilon(\eta):=(\epsilon_1(\eta)\dots \epsilon_n(\eta))$ must have 
a different value on each such segment. Since 
$\nu_k\cdot a=\nu_k^1 x_1+\nu_k^2 y_1=x_1(\nu_k^2\eta+\nu_k^1)$ 
for the locus $Y_e$ (remember that $\eta=y_1/x_1$ for such a locus), 
we have $\epsilon_k(e)=sign(\nu_k\cdot a)=sign(x_1)\epsilon_k(\eta)$.
The difference between the 
two edges $e$ and $-e$ of $\Delta$ which lie above an edge $(P_k,P_{k+1})$ of 
$\Delta_M$ is specified by the sign of $x_1$. In particular, 
the vector $\nu_e$ of equation (\ref{nu_e}) has the form:
\be
\nu_e=sign(x_1)\nu_\epsilon~~,
\ee
where\footnote{We warn the reader that the notation $\nu_\epsilon$ was used with a different meaning in the paper \cite{toric}.} $\nu_\epsilon:=\sum_{k=1}^n{\epsilon_k(\eta)\nu_k}$. Since 
$gcd(\nu_e^1,\nu_e^2)=gcd(\nu_\epsilon^1,\nu_\epsilon^2)$, it follows
that the singularity type of the twistor space along $Y_e$ and $Y_{-e}$
can also be determined from the vector $\nu_\epsilon$. This observation 
will be used in Sections 4 and 5.

\paragraph{Observation 2} By using an $SL(2,\Z)$ transformation, one 
can always arrange that $\nu_j^2$ are non-vanishing for all $j$.
In this case, none of the vertices $\eta_j$ of $\Delta_M$ coincide with the 
compactification point $\eta=\infty$. Throughout this paper, we shall assume 
that the toric hyperkahler generators satisfy this condition. 
 
\paragraph{Observation 3} 
The coordinates $\psi,\phi$ appearing in (\ref{Mmetric}) may have to be 
subjected to a $\Z_2$ identification. Let us assume for simplicity that the $\Z_2$ subgroup $\{ -1, 1 \}$ of $\H^*$ acts nontrivially on $X$. Then $\psi,\phi$ are taken to have periodicity $2\pi$. When descending to $M$ via the $\H^*$ quotient, the action of the $\Z_2$ subgroup $\{ -1, 1 \}$ of $Sp(1)\subset \H^*$ projects trivially through ${\vec \pi}$, and therefore identifies the $T^2$ fiber of $X$ with itself. This means that $\psi,\phi$ give coordinates on a double cover of the fiber of $M$. Let us consider the degeneration of the latter above an edge $e$ of $\Delta_M$. As explained in \cite{toric}, this corresponds to the locus $X_e$ in $X$ defined by the equations $w_j^{(-\epsilon_j(e))} = 0$. From the results of \cite{toric}, we know that the fiber of $X$ is non-degenerate above this locus. However, a certain $U(1)$ subgroup of $Sp(1)\subset \H^*$ (given in (\ref{U1proj})) will act vertically above such an edge. This means that the restriction of ${\vec \pi}$ to $X_e$ is invariant not only under the $\Z_2$ action mentioned above, but under the action of this whole $U(1)$ subgroup. This is why the fiber of $M$ degenerates to a circle above this edge: the circle is the quotient of the $T^2$ fiber of $X$ through this $U(1)$ action. Because the $\Z_2$ subgroup lies inside $U(1)$, the identification mentioned above is automatically implemented by the $U(1)$ quotient, and the coordinate induced on the $S^1$ fiber of $M$ above $e$ will not be subject to any further discrete quotient. The situation is more complicated above a vertex $\eta_j$ of $\Delta_M$. As explained in \cite{toric}, the associated locus in $X$ is given by $X_j = \{ u \in X | u_j = 0\}$. In this case, the $T^2$ fiber of $X$ collapses to a circle on $X_j$, and the $U(1)$ subgroup of $Sp(1)$ does not fix its ${\vec \pi}$-projection --- this projection is fixed only by the $\Z_2$ subgroup. If the $\Z_2$ action is nontrivial when restricted to $X_j$, then the fiber of $M$ above the vertex $\eta_j$ is a $\Z_2$ quotient of the $S^1$ fiber of $X$ along $X_j$; this means that the coordinate induced by $\psi, \phi$ along the fiber of $M$ must be subjected to a $\Z_2$ identification. If the $\Z_2$ subgroup acts trivially on $X_j$, then such a further identification does not occur. Which of these cases occurs can be decided by using the criteria of \cite{toric}. A similar analysis goes through if the $\Z_2$ subgroup of $Sp(1)$ acts trivially on $X$, with the conclusion that a further identification must be present above the entire polygon $\Delta_M$. Due to such double cover issues, the gauge group content computed in Sections 4 and 5 below may have to be 'halved' in certain situations. In this paper, we shall make the following simplifying assumptions:

(1) First, we require that the $\Z_2$ subgroup of $Sp(1)$ acts nontrivially on $X$ (a criterion for when this happens can be found in \cite{toric}).

(2) We shall further assume that the fibers of $M$ are not subject to any further identifications above the vertices of the polygon $\Delta_M$ (this holds, for example, for the model discussed in Section 2 of \cite{toric}).

These assumptions will allow us to avoid a lengthy case by case analysis and streamline the presentation. The modifications needed in the general case can be easily recovered by using the results of \cite{toric}.

\section{Reduction to IIA}
\setcounter{equation}{0}

We are now ready to explore the $G_2$ metric 
(\ref{G2deformed}), where $d \sigma^2$ is given explicitly 
by (\ref{Mmetric}). Remember that the $G_2$ metric admits two commuting $U(1)$ 
isometries (given by shifts of the coordinates $\phi$ and $\psi$), 
and that the compact isometries they generate are in correspondence with the 
two-dimensional lattice spanned by the toric hyperkahler generators $\nu_j$. 
Hence we can write 
down the $10$-dimensional IIA background obtained by reduction 
along one of these directions. We can always choose $\phi$ without 
loss of generality; any other choice (i.e. any primitive, integral 
linear combination
of $\phi$ and $\psi$) can be reduced to this upon performing modular 
transformations of the lattice spanned by $\nu_1\dots \nu_n$ 
(this amounts to changing the integral basis in 
which these vectors are expressed). With our notations, the coordinates 
$\psi,\phi$ correspond respectively 
to the first and second coordinates on $\Z^2$. 
Therefore, the vector $\nu\in \Z^2$ corresponds to
an isometry in the direction $\phi$ if and only if  $\nu^1=0$.

\subsection{The reduction}

The reduction along $\phi$ results from 
the standard KK ansatz: 
\be ds^2_{(11)} =
e^{-\frac{2}{3} \varphi_A (x)} g_{\mu \nu}(x) dx^{\mu} dx^{\nu} +
e^{\frac{4}{3} \varphi_A (x)} (d \phi + dx^{\mu} C_{\mu}(x))^2 \, .
\label{anzats} 
\ee 
This produces a $10$-dimensional IIA bosonic background
$(\varphi_A, g_{\mu \nu}, B_{\mu \nu} = 0)$ in the NS-NS sector and
$(C_{\mu}, A_{\mu \nu \rho} = 0)$ in the RR sector. In our case the
$B$ field and the three form $A_{\mu \nu \rho}$ vanish because we
start with a vacuum configuration in $11$ dimensions: \be ds^2 =
ds^2 ({\bf E}^{3,1}) + \frac{1}{f} dr^2 + \frac{r^2}{2}[ d \sigma^2 +
\frac{f}{2} |d_A {\vec u} |^2 ] \label{11metric} \ee and vanishing 
$11d$ supergravity three-form field.  In (\ref{11metric}) ${\bf E}^{3,1}$ is
$4d$ Minkowski space and $f(r) = 1 - (r_0/r)^4$.

Upon defining:
{\footnotesize\bea
\label{ABDE}
A(\rho , \eta) &=& \frac{ F^2 - 4\rho^2(F_\rho^2 + F_\eta^2) }{4 F^2 \rho^2} \nn \\
B(\rho,\eta) &=& 
\frac{4\left(F_\eta^2+F_\rho^2 \right)\rho^3-4\rho^{2}F F_\rho + 
\left[F ^2+4\eta^2(F_\rho^2+F_\eta^2)-8\eta FF_\eta \right] \rho+
4\eta^2F F_\rho +\frac {\eta^2}{\rho} F^2}{ F^2\left[F^2-4\rho^2(F_\rho^2 + F_\eta^2)\right]}\nn\\
D(\rho,\eta) &=& \frac{4\left( F_\eta^2+F_\rho^2 \right) \rho+
4F F_\rho +\frac {F^2}{\rho}}{ F^2\left[F^2-4\rho^2(F_\rho^2 + F_\eta^2)\right]}~~,\\
E(\rho,\eta) &=&
\frac{4\left[ -F_\eta F  +\eta(F_\eta^2+F_\rho^2) \right]\rho+
4\eta F F_\rho +\frac {\eta}{\rho} F^2}{ F^2\left[F^2-4\rho^2(F_\rho^2 + F_\eta^2)\right]}~~,\nn
\eea}
we can write the Einstein self-dual metric on $M$ in the following form:
\be d \sigma^2 = A (d \rho^2 + d \eta^2) + B \, d \phi^2 + D \, d
\psi^2 + 2 \, E \, d \phi \, d \psi \, . \label{esd} \ee Writing the connection
on the bundle $\Lambda^{2,-} (T^* M)$ as $A^j = A^j_{\rho} (\rho ,
\eta) d \rho + A^j_{\eta} (\rho , \eta) d \eta + A^j_{\phi} (\rho ,
\eta) d \phi + A^j_{\psi} (\rho , \eta) d \psi$\footnote{The
components of $A$ do not depend on $\phi$ and $\psi$ because the 
isometries of $M$ lift to isometries of
the twistor space, as discussed in Section 2 (below eq. (\ref{G2deformed})).}, $j=1,2,3$ and $\epsilon^{ijk} A^j u^k$ as
$(\vec{A} \times \vec{u})^i$, 
we can split $|d_A {\vec u} |^2=(du^i + \epsilon^{ijk} A^j u^k)^2$ into the
following terms: \be |d_A {\vec u} |^2 = H({\vec u}, \rho ,\eta) + P_{\phi}
\, d \phi^2 + P_{\psi} \, d \psi^2 + 2 \, P_{\phi \psi} \, d \phi \, d
\psi + 2 \, Q_{\phi} \, d \phi + 2 \, Q_{\psi} \, d \psi \,\, . \ee 
Here:
\be
\!\!\!H = \left[d{\vec u} +(\vec{A}_{\rho} \times \vec{u}) \, d\rho + 
(\vec{A}_{\eta} \times \vec{u}) \, d\eta \right]^2
\ee
is a 2-form independent of $\phi$ and $\psi$ and:
\bea
\label{HPQ}
P_{\phi} &=& |\vec{A}_{\phi} \times \vec{u}|^2 \nn\\ 
P_{\psi} &=& |\vec{A}_{\psi} \times \vec{u}|^2 \nn \\ 
P_{\phi \psi}&=& 
(\vec{A}_{\phi} \times \vec{u}) \, \cdot (\vec{A}_{\psi} \times \vec{u}) \\ 
Q_{\phi} &=& d\vec{u} \, \cdot \, (\vec{A}_{\phi} \times
\vec{u}) + (\vec{A}_{\rho} \times \vec{u}) \, \cdot (\vec{A}_{\phi} \times
\vec{u}) \, d \rho + (\vec{A}_{\eta} \times \vec{u}) \, \cdot (\vec{A}_{\phi}
\times \vec{u}) \, d \eta  \nn \\ Q_{\psi} &=&  d\vec{u} \, \cdot \,
(\vec{A}_{\psi} \times \vec{u}) + (\vec{A}_{\rho} \times
\vec{u}) \, \cdot (\vec{A}_{\psi} \times \vec{u}) \, d \rho + (\vec{A}_{\eta}
\times \vec{u}) \, \cdot  (\vec{A}_{\psi} \times \vec{u}) \, d \eta \,\, .\nn
\eea 
Recall that $u^i$ satisfy the constraint $\sum_{i=1}^3{u^iu^i}
= 1$. The connection $A = A^1 \i + A^2 \j + A^3 \k$ was calculated in
Section 8 of \cite{CP}: \bea
\label{A}
&&\vec{A}_{\phi} = \left( 0 \, ,\, -\frac{\sqrt{\rho}}{F} \, , \,
\frac{\eta}{F \sqrt{\rho}} \right) \qquad \,\,\, \vec{A}_{\psi} = \left( 0 \, , \,
0 \, , \, \frac{1}{F \sqrt{\rho}} \right) \nn \\ &&\vec{A}_{\rho} =
\left( -\frac{F_{\eta}}{F} \, ,\, 0 \, , \, 0 \right) \qquad \,\,\, ~~~~~~~\vec{A}_{\eta}
= \left( \frac{1}{2 \rho} + \frac{F_{\rho}}{F} \, , \, 0 \, , \, 0 \right) \, .
\eea Using this result, we obtain: 
\bea
H = |d{\vec u}|^2 &+& [1-(u^1)^2] \left[ \frac{F_{\eta}}{F} d\rho -
\left( \frac{1}{2 \rho} + \frac{F_{\rho}}{F} \right) d \eta \right]^2 +
\\ &+& 2 \left[
\frac{F_{\eta}}{F} d\rho - \left( \frac{1}{2 \rho} + \frac{F_{\rho}}{F} \right) d
\eta \right] (u^3 du^2 - u^2 du^3)\nn
\eea
and:
\bea 
\label{PQ}
P_{\phi} &=& \frac{(u^2 \eta +
u^3 \rho)^2 + (u^1)^2 (\rho^2 + \eta^2)}{\rho F^2} = \frac{\rho^2 +
\eta^2 - (u^2 \rho - u^3 \eta)^2}{\rho F^2} \nn \\ 
P_{\psi} &=&
\frac{(u^1)^2 + (u^2)^2}{\rho F^2} =\frac{1-(u^3)^2}{\rho F^2}\nn \\ 
P_{\phi \psi} &=&
\frac{[(u^1)^2 + (u^2)^2] \eta + u^2 u^3 \rho}{\rho F^2} = 
\frac{\eta
+ u^3 (u^2 \rho - u^3 \eta)}{\rho F^2} \nn \\ 
Q_{\phi} &=& \frac{1}{\sqrt{\rho} F} \Bigg\{
u^1 (\eta \, du^2 + \rho \, du^3) - (u^2 \eta + u^3 \rho) \, du^1 \Bigg. \nn \\
&+& \left. u^1 (u^3 \eta - u^2 \rho) \left[ \frac{F_{\eta}}{F} d\rho 
- \left( \frac{1}{2 \rho} + \frac{F_{\rho}}{F} \right) d \eta \right] \right\} \nn \\
Q_{\psi} &=& \frac{1}{\sqrt{\rho} F} \left\{ u^1 du^2 - u^2 du^1 + u^1 u^3 \left[ \frac{F_{\eta}}{F} d\rho - \left( \frac{1}{2 \rho} + \frac{F_{\rho}}{F} \right) d \eta \right] \right\} \, .  \eea 
The second equalities in $P_{\phi}$, $P_\psi$  and $P_{\phi
\psi}$ are obtained upon using the constraint $|{\vec u}|^2 = 1$.
Let us define a symmetric two by two matrix $U$ with entries:
\bea
U_{11} = B + \frac{f}{2} P_{\phi}~~,~~U_{22}= D + \frac{f}{2} P_{\psi}~~,~~
U_{12}=U_{21} = E + \frac{f}{2} P_{\phi \psi}~~.
\eea
This allows us to write the eleven-dimensional metric (\ref{11metric}) in the form: 
\be
ds^2 = ds^2 ({\bf E}^{3,1}) + \frac{1}{f} dr^2 + 
\frac{r^2}{2}[U_{11} \, d \phi^2 + U_{22}\, d \psi^2 + 2 \, U_{12} \, d \phi \, d \psi + 2 \, \bar{Q}_{\phi} \, d \phi + 2 \, \bar{Q}_{\psi} \, d \psi + \bar{H}]~~,
\ee
where:
\be
\bar{H} = \frac{f}{2} H + A \, (d \rho^2 + d \eta^2)~~ 
\ee
and 
\be
\label{Qbar}
\bar{Q}_{\phi} = \frac{f}{2} Q_{\phi}~~,~~
\bar{Q}_{\psi} = \frac{f}{2} Q_{\psi} 
\ee
To further simplify this expression, we introduce the one-forms:
\be
\alpha_1 = \frac{U_{22} \bar{Q}_{\phi} - U_{12} \bar{Q}_{\psi}}{det U} \, , 
\qquad \alpha_2 = \frac{U_{11} \bar{Q}_{\psi} - 
U_{12} \bar{Q}_{\phi}}{det U}~~ 
\label{forms}
\ee
as well as the two-form:
\be
\bar{h} = \bar{H} + U_{11}\alpha_1^2 + 2 U_{12} \alpha_1 \alpha_2 +U_{22} \alpha_2^2~~,
\ee
all of which depend only on $\rho,\eta$ and ${\vec u}$.
If we let $\phi_1 \equiv \phi$ and $\phi_2 \equiv \psi$, then the 11d metric 
becomes: 
\be ds^2 = ds^2 ({\bf
E}^{3,1}) + \frac{1}{f} dr^2 + \frac{r^2}{2}[U_{ij} \, (d\phi_i + \alpha_i) \, 
(d\phi_j + \alpha_j) + \bar{h}] \, .
\ee  
The virtue of this form of the metric is that it is adapted to the $T^2$ fibration of the ESD base 
$M$ over the hyperbolic plane $(\rho,\eta)$, 
which in turn induces a similar fibration of the $G_2$ cone 
over a five-dimensional space spanned by the coordinates 
$r, \rho, \eta$ and ${\vec u}$.
The entries of the matrix $U$ characterize the metric induced 
on the $T^2$ fibers.

Recall that we want to perform dimensional reduction along
$\phi \equiv \phi_1$. Comparing the last equation with (\ref{anzats}) we read off the IIA dilaton and RR 1-form: 
\be
\label{dilaton_C}
\varphi_A = \frac{3}{4} \ln \bigl(\frac{r^2 U_{11}}{2} \bigr) \qquad C =
\frac{U_{12} d \psi + \bar{Q}_{\phi}}{U_{11}} = \frac{U_{12} d \psi + \alpha_1 U_{11} + \alpha_2 U_{12}}{U_{11}} \, .  \label{phiC} \ee For the IIA
metric in string frame we obtain: 
\bea 
\label{metricA}
ds_{A}^2 
&=& \left(\frac{r^2 U_{11}}{2}\right)^{1/2}  
\{ ds^2 ({\bf E}^{3,1}) + \frac{1}{f} dr^2 +
\frac{r^2}{2 U_{11}} [ \det U \, d \psi^2 + 
2 \, (U_{11} \bar{Q}_{\psi} - U_{12} \bar{Q}_{\phi}) \, d \psi \nn \\
&-& \bar{Q}_{\phi}^2 +  
U_{11} \bar{H} ] \} \nn \\
&=& \left(\frac{r^2 U_{11}}{2}\right)^{1/2}  
\{ ds^2 ({\bf E}^{3,1}) + \frac{1}{f} dr^2 +
\frac{r^2}{2 U_{11}} [ \det U \, d \psi^2 + 2 \, \det U \, 
\alpha_2 \, d \psi \nn \\
&-& 2 \, (\alpha_1 U_{11} + \alpha_2 U_{12})^2 - \alpha_2^2 \, \det U +
U_{11} \bar{h} ] \}
~~, \label{IIA} \eea
\noindent which in particular gives the components: 
\bea g^{(A)}_{\psi \psi} &=&
\left(\frac{r^2U_{11}}{2}\right)^{1/2}
\frac{r^2 \, \det U}{2 \, U_{11}} \nn \\ 
g^{(A)}_{\psi \rho} &=&
\left(\frac{r^2 U_{11}}{2}\right)^{1/2}
\frac{r^2 f u^1}{4 U_{11}F \sqrt{\rho} }\frac{F_\eta}{F}
[U_{11} u^3 - U_{12} (u^3 \eta - u^2 \rho)] \nn \\ 
g^{(A)}_{\psi \eta} &=&-
\left(\frac{r^2U_{11}}{2}\right)^{1/2}
\frac{r^2 f u^1}{4 U_{11} F \sqrt{\rho}} \left( \frac{1}{2 \rho}
+ \frac{F_{\rho}}{F} \right) [U_{11}u^3-U_{12}(u^3 \eta - u^2 \rho )] \\
g^{(A)}_{\psi u^1} &=& \left(\frac{r^2U_{11}}{2}\right)^{1/2}
\frac{r^2 f}{4 U_{11} F\sqrt{\rho}}
\left[ U_{12} \left( u^2 \eta + \frac{1-(u^2)^2}{u^3}\rho \right) - U_{11}u^2 \right]
\nn \\ 
g^{(A)}_{\psi u^2} &=& -\left(\frac{r^2U_{11}}{2}\right)^{1/2}
\frac{r^2 f}{4 U_{11} F\sqrt{\rho}} \left[ U_{12} \left(u^1 \eta - \frac{u^1 u^2\rho}{u^3} \right)-U_{11}u^1 \right]~~,\nn
\eea
where we used the relation $du^3 = - (u^1 \, du^1 + u^2 \,
du^2)/u^3$. The remaining $g_{\psi m}$ are zero.

Let us introduce spherical coordinates
\footnote{The reader may wonder why we didn't do this from the beginning. 
One reason is that the expressions in terms of $u$ are more compact. For some
purposes they are more convenient, for others the spherical
coordinates are preferable. We note that the most symmetric form of
the metric components is attained if one sets $u^1 = \cos \theta$ and
$u^2 = \sin \theta \cos \chi$. However it will be more convenient
later to use the definitions in (\ref{sphcoord}).}$\theta,\chi$ through the 
relations:
\be 
u^3=\cos\theta \, , \qquad u^1=\sin\theta \cos\chi \, , \qquad 
u^2=\sin\theta \sin \chi \, , \label{sphcoord}
\ee
where $\theta\in [0,\pi]$ and $\chi\in [0,2\pi]$.
One computes: 
\bea
g_{\psi \chi}^{(A)} &=& \left(\frac{r^2U_{11}}{2}\right)^{1/2}
\frac{r^2 f}{4 U_{11} F\sqrt{\rho}} 
\left[ U_{11} \sin^2 (\theta) - U_{12} \left(\frac{\rho}{2} 
\sin (2\theta) \sin \chi + \eta \sin^2 (\theta) \right) \right] \nn \\
g_{\psi \theta}^{(A)} &=& \left(\frac{r^2U_{11}}{2}\right)^{1/2}
\frac{r^2 f}{
4 U_{11} F} U_{12} \sqrt{\rho} \cos \chi~~\\
g_{\theta \theta}^{(A)} &=& \left(\frac{r^2U_{11}}{2}\right)^{1/2}
\left[
\frac{r^2 f}{4} - \frac{r^2 f^2}{8 U_{11} \rho F^2} \rho^2 \cos^2 (\chi) \right] \
\nn \\
g_{\chi \chi}^{(A)} &=& \left(\frac{r^2U_{11}}{2}\right)^{1/2}
\left[\frac{r^2 f}{4} \sin^2 (\theta) - \frac{r^2 f^2}{8 U_{11} \rho F^2} 
\sin^2 (\theta) (\eta \sin \theta + \rho \cos \theta \sin \chi)^2 \right]\nn \\
g_{\chi \theta}^{(A)} &=& \left(\frac{r^2U_{11}}{2}\right)^{1/2}
\frac{r^2 f^2}{8 U_{11} \rho F^2} 2 \rho \sin \theta \cos 
\chi (\rho \cos \theta \sin \chi + \eta \sin \theta)~~.\nn
\eea
In these relations, we have isolated the conformal factor 
$\left(\frac{r^2U_{11}}{2}\right)^{1/2}$ which multiplies the metric
(\ref{metricA}). Since we will be interested in comparing internal sizes
with sizes in the Minkowski directions ${\bf E}^{3,1}$, 
it will often be convenient 
to work with metric components ${\tilde g}_{ij}$ which do not contain
this factor:
\be
g_{ij}=\left(\frac{r^2U_{11}}{2}\right)^{1/2}{\tilde g}_{ij}~~.
\ee
All distances computed with respect to ${\tilde g}_{ij}$
will be indicated by a tilde. 
In particular, we shall be interested in the 
metric induced by ${\tilde g}_{ij}$ 
on the $S^2$ fiber of $Y \rightarrow M$:
\be 
\label{S2metric}
d{\tilde s}^2|_{\theta,\chi} = \frac{r^2}{2} \left[ \frac{f}{2}|d\vec{u}|^2 -
\frac{\bar{Q}_{\phi}^2}{U_{11}}|_{du^i du^j} \right] \, . \label{fiber}\ee
Also note that the square radius of the circle spanned by $\psi$ 
(with respect to the modified metric ${\tilde g}_{ij}$)  equals 
${\tilde R}_\psi^2={\tilde g}_{\psi\psi}=
\frac{r^2}{2}\frac{\det U}{U_{11}}$.
Finally, we list the 
nonzero components of the RR one-form, which can be read off from (\ref{phiC}):
\bea
C_{\psi} &=& \frac{U_{12}}{U_{11}} \nn \\
C_{\rho} &=& \frac{f(r) \, u^1 F_{\eta}}{2 \, U_{11} \, \sqrt{\rho} \, F^2} (u^3 \eta -u^2 \rho) \nn \\
C_{\eta} &=& \frac{f(r) \, u^1}{2 \, U_{11} \, \sqrt{\rho} \, F} (u^2 \rho - u^3 \eta) \left(\frac{1}{2 \rho} + \frac{F_{\rho}}{F}\right)  \\
C_{u^1} &=& - \frac{f(r)}{2 \, U_{11} \, \sqrt{\rho} \, F} \left[u^2 \eta + u^3 \rho + \frac{(u^1)^2 \rho}{u^3} \right] \nn \\
C_{u^2} &=& \frac{f(r) \, u^1}{2 \, U_{11} \, \sqrt{\rho} \, F} \left[ \eta - \frac{u^2 \rho}{u^3} \right]~~.\nn
\eea

\subsection{Behavior of the IIA solution for $\rho\rightarrow 0$}
\setcounter{equation}{0}

As we saw above, the singularities of our $G_2$ spaces occur above
the boundary $\rho = 0$ of ${\overline {\cal H}}^2$. Following
\cite{AW, Witten_Acharya}, we expect that the dimensional reduction will 
give a
type IIA D-brane configuration and that the D-branes will be localized
precisely when the reduction is performed through an isometry of the
$11$-dimensional metric which fixes the singular loci.\footnote{The prototype example of this construction is the relation between the M-theory Kaluza-Klein monopole and the flat type IIA D6-brane \cite{Townsend}.} We will confirm
this expectation by a careful study of the asymptotics of the
$10$-dimensional fields in the limit $\rho
\rightarrow 0$. Before embarking on the analysis of the horizontal and
vertical loci described in Subsection 2.2, we need some preparatory calculations of 
the asymptotics of various expressions.

Following \cite{toric}, we shall assume that the toric hyperkahler cone $X$ 
is {\em good}, which means that any two of the toric hyperkahler generators
are linearly independent. This assumption will be made for the remainder of 
this paper. 

From (\ref{F}) one can see that the behavior of 
$F$ and $F_\eta$ for $\rho\rightarrow 0$ is controlled at all orders by the 
quantities: 
\be
L_j(\eta):=\sum_{k=1}^n\frac{\epsilon_k(\eta)^j
(\nu_k^2)^j}{|\nu_k^2\eta+\nu_k^1|^{j-1}}~~~~~~~(j\geq 0)~~,
\ee
where $\epsilon_k(\eta):=sign(\nu_k^2\eta+\nu_k^1)$.
The leading orders have the form:
{\footnotesize \be
\label{F_as}
F=L_0(\eta)\rho^{-1/2}+\frac{1}{2}L_2(\eta)\rho^{3/2}
-\frac{1}{8}L_4\rho^{7/2}+O(\rho^{11/2})~,~
F_\eta=L_1\rho^{-1/2}
-\frac{1}{2}L_3\rho^{3/2}+\frac{3}{8}L_5\rho^{7/2}+O(\rho^{11/2})~~.
\ee}
We note that $L_1(\eta)$ coincides with $\frac{d}{d\eta}L_0(\eta)$, 
and that:
\be
L_0=\sum_{k=1}^n{|\nu_k^2\eta+\nu_k^1|}~~,~~
L_1=\sum_{k=1}^n{\epsilon_k(\eta)\nu_k^2}=\nu_{\epsilon(\eta)}^2~~,~~
L_2=\sum_{k=1}^n{\frac{(\nu_k^2)^2}{|\nu_k^2\eta+\nu_k^1|}}~~,
\ee
where $\epsilon(\eta)=(\epsilon_1(\eta)\dots \epsilon_n(\eta))$ and
$\nu_\epsilon=\sum_{k=1}^n{\epsilon_k(\eta)\nu_k}$ (See Section 3). 
With our notations, one has:
\be
\nu_\epsilon=\left[\begin{array}{c}T\\L_1\end{array}\right]~~.
\ee
In particular, $T$ and $L_1$ are constant on each edge of $\Delta_M$, 
but each of them will generally jump when one passes from one edge 
to another. 

Upon noticing that $|\nu_k^2\eta+\nu_k^1|=
\epsilon_k(\eta)(\nu_k^2\eta+\nu_k^1)$, we obtain:
\be
L_0(\eta)=\sum_{k=1}^n{\epsilon_k(\eta)(\nu_k^2\eta+\nu_k^1)}=
\eta L_1(\eta)+T(\eta)~~,
\ee
with $T(\eta)=\sum_{k=1}^n{\epsilon_k(\eta)\nu_k^1}=
\nu_{\epsilon(\eta)}^1$. Note that $L_0$ is bounded from below by a strictly 
positive constant
depending on the toric hyperkahler generators $\nu_j$\footnote{ 
Indeed, we have $n\geq 2$ and the vectors $\nu_j$ cannot all be proportional
(remember that they have to span $\R^2$). This shows that $L_0(\eta)$ is
strictly positive on the real axis.  The statement then follows by
continuity of $L_0$ together with the fact that $L_0(\eta)$ tends to
$+\infty$ for $\eta\rightarrow \pm \infty$.  The lower bound is given
by $min_{\eta\in \R}L_0(\eta)>0$, which is attained for some value of
$\eta$. }, while $L_1$ and $T$ can be positive or negative and are
bounded both from below and from above. On the other hand, $L_2$ is
strictly positive for finite $\eta$ but vanishes in the limit
$\eta\rightarrow \pm\infty$. The latter function tends to $+\infty$ at
the values $\eta=\eta_j:=-\frac{\nu_j^1}{\nu_j^2}$.

\paragraph{Observation 1} In what follows, we shall encounter the 
quantity $\delta=L_0L_2-L_1^2$. It is important to realize that it is 
positive. To see this, notice that substituting 
$\epsilon_j(\eta)=\frac{\nu_j^2\eta+\nu_j^1}{|\nu_j^2\eta+\nu_j^1|}$
in the expression for $L_1$ gives:
\be
L_1=\eta L_2+S~~,
\ee
where $S=\sum_{j=1}^n{\frac{\nu_j^1\nu_j^2}{|\nu_j^2\eta+\nu_j^1|}}$. 
Therefore
\be
L_0=\eta^2L_2+\eta S+T \, .
\ee
Using the last two relations one obtains:
\be
\delta=L_0L_2-L_1^2=L_2(T-\eta S)-S^2~~.
\ee
Upon noticing that 
$T-\eta S=\sum_{j=1}^{n}{\frac{(\nu_j^1)^2}{|\nu_j^2\eta+\nu_j^1|}}$, 
this leads to the expression:
\be
\delta=\left(\sum_{j=1}^n{\frac{(\nu_j^1)^2}{|\nu_j^2\eta+\nu_j^1|}}\right)
\left(\sum_{j=1}^n{\frac{(\nu_j^2)^2}{|\nu_j^2\eta+\nu_j^1|}}\right)-
\left(\sum_{j=1}^n{\frac{\nu_j^1\nu_j^2}{|\nu_j^2\eta+\nu_j^1|}}\right)^2~~,
\ee
which is non-negative by the Schwartz inequality. 
It is clear that $\delta $ 
can be zero only if the two rows $(\nu^1_1\dots \nu^1_n)$ and 
$(\nu^2_1\dots \nu^2_n)$ of the  
matrix $G$ (see Subsection 2.1) are proportional; this is impossible 
since $G$ has maximal rank (recall that its columns $\nu_j$ must generate 
$\Z^2$). We conclude that $\delta $ is 
strictly positive. Note, however, that $\delta $ tends to zero for 
$\eta\rightarrow \pm \infty$.

\paragraph{Observation 2} The quantities 
$T$ and $\eta_j$ obey certain properties which will be used below.
First, notice that $T(0)=\sum_{k=1}^n{sign(\nu_k^1)\nu_k^1}=
\sum_{k=1}^n{|\nu_k^1|}$. Thus vanishing of $T(0)$ would require that 
all $\nu_k^1$ vanish, which is impossible since they must generate $\Z^2$. 
It follows that $T(0)$ cannot vanish. If $T$ vanishes along a given edge of 
$\Delta$, this clearly implies that $\eta=0$ cannot be an endpoint of this 
edge. Now suppose that we are 
given a model for which $\eta=0$ is a vertex of $\Delta_M$. Then the 
previous observation shows that $T$ must be non-vanishing along 
the two edges of $\Delta_M$ which share this vertex.
We also note that at most one of the toric hyperkahler generators can 
have the property $\nu_j^1=0$. Indeed, if $\nu^1_j=\nu^1_k=0$ for 
some $j\neq k$, then the generators $\nu_j$ and $\nu_k$ must be proportional, 
which contradicts the assumption that the toric hyperkahler cone $X$ is good.
As discussed in \cite{toric}, the requirement that $X$ is good means that
the principal diagonals of $\Delta$  cannot coincide, which means that 
$\Delta$ has $2n$ distinct vertices. This obviously implies that $\Delta_M$ 
must have $n$ (distinct) vertices, which is another explanation for our 
second observation above. We shall often use these basic properties in the 
analysis of Sections 4 and 5.

\subsubsection{Asymptotics of $U$}

Using (\ref{F_as}), one finds the following 
expansions for $\rho\rightarrow 0$:
\bea
\label{ABDE_as}
A&=&\frac{\delta}{L_0^2}+O(\rho^2)~~~,~~B=\frac{T^2}{L_0^2\delta}+
O(\rho^2)~~\nn\\
D&=&\frac{L_1^2}{L_0^2 \delta }+O(\rho^2)~~,~~
E=-\frac{L_1T}{L_0^2\delta}+O(\rho^2)
\eea
as well as:
\bea
\label{Uas}
U_{11}&=&
\frac{T^2}{L_0^2\delta}+\frac{f}{2}\frac{\eta^2}{L_0^2}\sin^2(\theta)
+O(\rho)~~\nn\\
U_{22}&=&\frac{L_1^2}{L_0^2 \delta}+
\frac{f}{2}\frac{1}{L_0^2}\sin^2(\theta)+O(\rho^2)~~\nn\\
U_{12}&=&-\frac{L_1T}{L_0^2\delta}+
\frac{f}{2}\frac{\eta}{L_0^2}\sin^2(\theta)+O(\rho)~~.
\eea
To arrive at the last equations, we used the coordinate transformation (\ref{sphcoord}) and the leading asymptotics 
of the connection 
(\ref{A}):
\bea
\label{Aas}
{\vec A}_\rho&=&\left(-\frac{L_1}{L_0},0,0 \right)+O(\rho^2)~~,
~~{\vec A}_\psi=\left(0,0,\frac{1}{L_0}\right)~~~+~O(\rho^2)~~\nn\\
{\vec A}_\phi&=&\left(0,0,\frac{\eta}{L_0}\right)~~+O(\rho)~~,
~~{\vec A}_\eta=\rho \, \left(\frac{L_2}{L_0},0,0 \right)+O(\rho^3)~~.
\eea

We shall also need the asymptotics of $det U$. Upon using the higher order asymptotics of ${\vec A}$ and $B,D,E$, one finds:
\be
\label{detUas}
det U=\frac{f}{2L_0^2}\frac{\sin^2(\theta)}{\delta}+
\frac{fL_1}{2L_0^3}\frac{\sin(2\theta)\sin\chi}{\delta}\rho +
\frac{\Lambda}{4L_0^4\delta^2}\rho^2+O(\rho^3)~~,
\ee
where:
\bea
\Lambda&=&f\delta(f\delta+2L_1^2)\sin^2(\theta)\cos^2(\chi)+
fL_0\left[L_2(5\delta-L_1^2)+L_0(2L_1L_3-L_0L_4)\right]\cos^2(\theta)+\nn\\
&&+4\delta^2+f\left[L_0^2(L_0L_4+L_2^2-2L_1L_3)-2\delta(5\delta+L_1^2)
\right]~~,
\eea 
with $\delta=L_0L_2-L_2^1$ as above.

\subsubsection{Asymptotics for the metric components along $\psi,\theta$ 
and $\chi$}

\begin{figure}[hbtp]
\begin{center}
\scalebox{0.6}{\input{ctorus.pstex_t}}
\end{center}
\caption{\label{ctorus} The two-torus spanned by $\psi$ and $\chi$. 
The upper figure is purely topological and 
shows two elementary cells in the associated integral lattice 
(whose generators are normalized to have length $2\pi$), 
for the case $T>0$. The filled-in cell corresponds to the cycles $\Gamma_\psi$ 
and $\Gamma_\chi$ (defined respectively by $\chi=0$ and $\psi=0$), 
while the rectangular cell on the left corresponds to $\Gamma_\xi$ 
and $\Gamma_\zeta$ (defined respectively by $\zeta=0$ and $\xi=0$). Taking 
both $(\psi,\chi)$ and $(\xi,\zeta)$ to belong to $[0,2\pi]^2$ gives good 
parameterizations of the same torus. The lower figure shows the metric 
geometry as $\rho$ approaches zero. In this limit, the length 
of $\Gamma_\zeta$ tends to zero, so that $\Gamma_\chi$ 
becomes a $T$-fold cover of $\Gamma_\xi$ (figure \ref{cycle}).}
\end{figure}
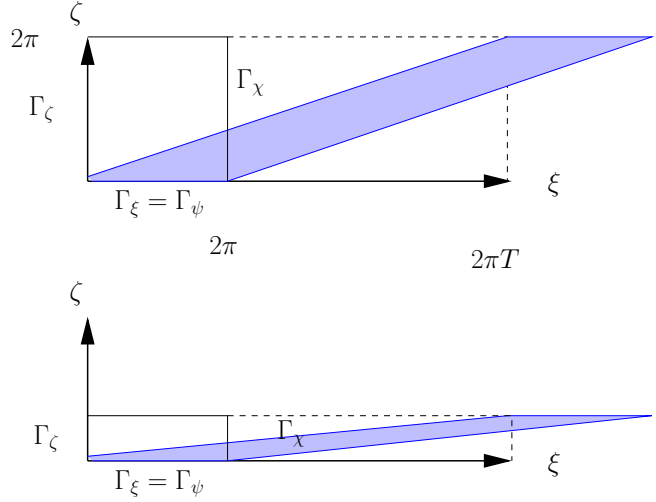

\begin{figure}[hbtp]
\begin{center}
\scalebox{0.4}{\input{cycle.pstex_t}}
\end{center}
\caption{\label{cycle} The cycle $\Gamma_\chi$ for the case $T=2$.}
\end{figure}
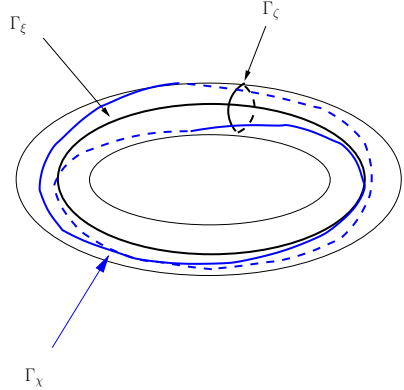

Consider the asymptotics of the angular metric components in 
the coordinates $\theta$, $\chi$ and $\psi$:
\bea
\label{angularas}
\tilde{g}_{\theta \theta}^{(A)} &=& \frac{r^2 f}{4} + O(\rho^2) \nn \\
\tilde{g}_{\theta \chi}^{(A)} &=& O(\rho) \nn \\
\tilde{g}_{\chi \chi}^{(A)} &=& \frac{r^2 f}{4} \frac{2T^2 \sin^2 
(\theta)}{2T^2 + f\eta^2 \delta \sin^2 (\theta)} + O(\rho)\nn \\
\tilde{g}_{\psi \chi}^{(A)} &=& 
\frac{r^2 f}{4} \frac{2 T \sin^2 (\theta)}{2T^2 + 
f \eta^2 \delta \sin^2 (\theta)} + O(\rho)\nn  \\
\tilde{g}_{\psi \theta}^{(A)} &=& O(\rho) \nn \\
\tilde{g}_{\psi \psi}^{(A)} &=& \frac{r^2 f}{4} 
\frac{2\sin^2 (\theta)}{2T^2 + f \eta^2 \delta \sin^2 (\theta)} 
+ O(\rho) \, .
\eea
The metric (\ref{S2metric}) induced on
the $S^2$ fiber takes the  following form in the limit $\rho \rightarrow 0$:
\be
\label{S2as}
d{\tilde s}^2|_{\theta,\chi} =
\frac{r^2 f}{4} \{ d\theta^2 + {\cal U}(r, \eta,\theta)
\sin^2 (\theta)  d\chi^2 \}+O(\rho)~~.
\ee
The squashing factor ${\cal U}$ is given by:
\be
\label{sqas}
{\cal U}(r,\eta,\theta) = \frac{2 T^2}{2 T^2 + 
f \eta^2 \delta \sin^2 (\theta)} + O(\rho) ~~.
\ee
However (\ref{angularas}) shows something more interesting, namely that
the metric induced in the directions $\theta,\psi$ and $\chi$ reduces to:
{\footnotesize \bea  
\label{induced_metric}
d{\tilde s}^2|_{\theta,\chi,\psi}= \frac{r^2 f}{4} \left[d\theta^2+
\frac{2\sin^2(\theta)}{2T^2 + f \eta^2\delta\sin^2(\theta)} (T(\eta)d\chi+d\psi)^2\right] +~~~O(\rho) ~~.
\eea}
At a fixed value of $\eta$, this shows that only one of the two periodic 
directions spanned by $\chi$ and $\psi$ has finite size in the limit 
$\rho\rightarrow 0$. Above a fixed edge of $\Delta_M$, one has 
$T=\nu_\epsilon^1$, and (\ref{induced_metric}) shows that the two-torus spanned
by $\chi$ and $\psi$ collapses to the cycle spanned by 
$\nu_\epsilon^1\chi+\psi$. To make this precise, consider the change of 
coordinates\footnote{We shall see in a moment that $\xi$ is a good 
coordinate in a whole tubular neighborhood of the locus $\rho=0$.}:
\be
\xi = T\chi+\psi~~,~~\zeta = \chi\Longleftrightarrow 
\psi = \xi-T\zeta~~,~~\chi = \zeta~~.
\ee
The two-torus spanned by $\chi$ and $\psi$ can be viewed as 
the quotient of $\R^2$ through the integral lattice generated by the 
tangent vectors $\partial_\chi$ and $\partial_\psi$. 
The transformation $(\psi,\chi)\rightarrow (\xi,\zeta)$
corresponds to an invertible 
integral transformation implemented by the modular 
matrix $S=\left[\begin{array}{cc}1&0\\-T&1
\end{array}\right]\in SL(2,\Z)$:
\be
\label{qmod}
\left[\begin{array}{c}\partial_\xi\\\partial_\zeta\end{array}\right]=S
\left[\begin{array}{c}\partial_\psi\\\partial_\chi\end{array}\right]~~,
\ee
i.e.:
\be
\partial_\xi=\partial_\psi~~,~~\partial_\zeta=\partial_\chi-T\partial_\psi
\Longleftrightarrow
\partial_\psi=\partial_\xi~~,~~\partial_\chi=T\partial_\xi+\partial_\zeta~~.
\ee
Since (\ref{qmod}) 
preserves the lattice generated by $\partial_\chi$ and $\partial_\psi$, 
we obtain new coordinates on this torus 
upon taking $\xi,\zeta$ to belong to $[0,2\pi]$. We have: 
\be
d\xi=Td\chi+d\psi~,~~d\zeta=d\chi\Longleftrightarrow  
d\psi=d\xi-Td\zeta~,~~d\chi=d\zeta~~.
\ee
The induced metric (\ref{induced_metric}) becomes:
\be
\label{im}
d{\tilde s}^2|_{\theta,\chi,\psi}=d{\tilde s}^2|_{\theta,\zeta}= 
\frac{r^2 f}{4} 
\left[d\theta^2+
\frac{2\sin^2(\theta)}{2T^2 + f \eta^2\delta 
\sin^2(\theta)} d\xi^2\right] + O(\rho) \nn~~.
\ee
In particular, one has:
\bea
{\tilde R}_\xi^2~~&=&||\partial_\xi||^2=\frac{r^2f}{4}
\frac{2\sin^2(\theta)}{2T^2 + f \eta^2 \delta\sin^2(\theta)}+
O(\rho)~~,\nn\\
{\tilde R}_\zeta^2~~&=&||\partial_\zeta||^2=O(\rho)~~\\
\partial_\xi\cdot \partial_\zeta&=&O(\rho)~~.\nn
\eea
Hence the $\xi$- and $\zeta$-cycles become orthogonal for small $\rho$, 
while the $\zeta$-cycle collapses to zero size (figure \ref{ctorus}). 
In this limit, the $\chi$-cycle becomes a $T$-fold cover of the $\xi$-cycle
(figure \ref{cycle}).

\noindent

\noindent

To understand the global situation above the polygon $\Delta_M$, 
consider the behavior of (\ref{induced_metric}) near a vertex 
$\eta=\eta_j:=-\frac{\nu_j^1}{\nu_j^2}$. 
Since $L_0$ remains strictly positive while 
$L_2(\eta)$ tends to infinity like $\frac{1}{|\eta-\eta_j|}$ 
for $\eta\rightarrow \eta_j$, the behavior of the quantity $\eta^2\delta$ 
in this limit depends on whether $\eta_j$ is nonzero or not, i.e. 
$\nu_j^1\neq0$ or $\nu_j^1=0$. 
If $\nu_j^1\neq 0$, we have $\eta^2\delta\rightarrow\infty$ and 
the second term in (\ref{induced_metric}) vanishes. This shows 
that (\ref{induced_metric}) is continuous when $\eta$ passes through the
value $\eta_j$ at $\rho=0$. 
If $\nu_j^1=0$, then $\eta^2\delta$ tends to zero for 
$\eta\rightarrow \eta_j=0$. On the other hand, vanishing of $\nu_j^1$ 
shows that the quantity 
$T(\eta)=\nu^1_{\epsilon(\eta)}$ reduces to 
$\sum_{k\neq j}{\epsilon_k(\eta)\nu_k^1}$, which has the same value on the 
two edges of $\Delta_M$ adjacent to the vertex $\eta_j$. Therefore, 
the coefficients of the 
induced metric (\ref{induced_metric}) are still continuous at $\eta=\eta_j$,
where the metric takes the form:
\be
\label{im_r}
d{\tilde s}^2|_{\theta,\chi,\psi}=\frac{r^2 f}{4} 
\left[d\theta^2+
\frac{\sin^2(\theta)}{T^2} (Td\chi+d\psi)^2\right] + O(\rho) \nn~~.
\ee
This shows that there exists a good coordinate $\xi$, 
defined on a tubular neighborhood of the locus $\rho=0$, such that
$\xi=T(\eta)\chi+\psi+O(\rho)$  for $\rho\rightarrow 0$. 
This allows us to write (\ref{induced_metric}) in the form (\ref{im})
with respect to this coordinate. 

Because the $\zeta$-cycle is always collapsed in our limit, 
the locus $\rho=0$ has codimension {\em two}. Fixing the radial coordinate 
$r$, this locus is described by a fibration over the edges 
of $\Delta_M$. 
If 
$T=\nu_\epsilon^1$ vanishes along some edge, then the 
induced metric (\ref{im}) reduces to:
\be
\label{im_cyl}
d{\tilde s}^2|_{\theta,\chi,\psi}= \frac{r^2 f}{4} \left[d\theta^2+
\frac{2}{f \eta^2\delta} d\xi^2\right] \nn~~.
\ee
Hence the $S^2$ fibers reduce to cylinders of (relative) 
height $\frac{\pi r\sqrt{f}}{2}$ and radius $\frac{r}{|\eta|\sqrt{2\delta}}$. 
The cylinder's radius collapses to zero when 
$\eta$ approaches the endpoints of this edge (none of which can lie at 
$\eta=0$). Therefore, the whole locus sitting above this edge at $\rho=0$ 
is a cylinder whose cross section is a two-sphere obtained by fibering 
the $\xi$-circle above this edge (the radius of the 
$\xi$-circle collapses above the endpoints). Also note that $\xi=\psi$ 
along this locus, so that the $\xi$ and $\psi$ circles coincide.

\begin{figure}[hbtp]
\begin{center}
\mbox{\epsfxsize=10truecm \epsffile{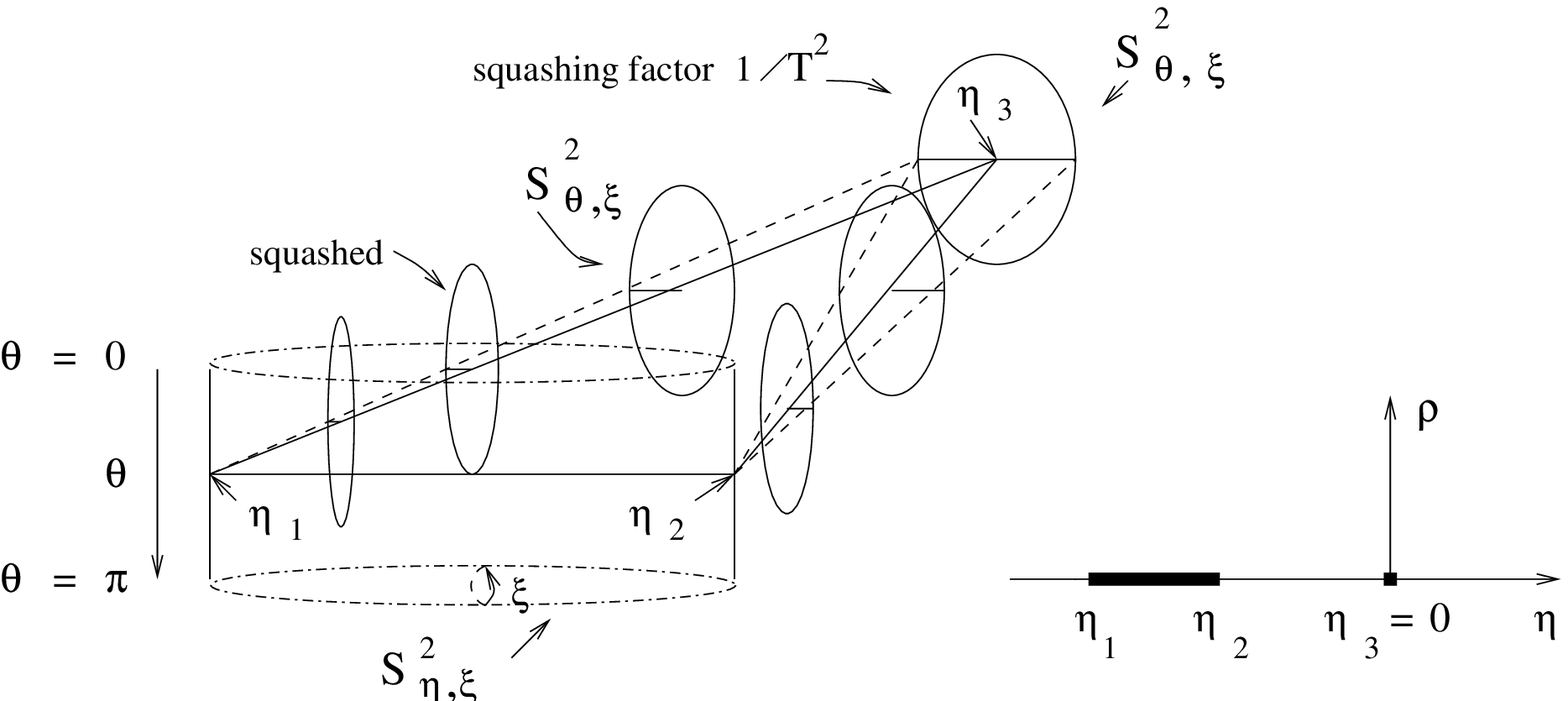}}
\end{center}
\caption{\label{2Ageometry} The locus $\rho=0$ at a fixed value of $r$. 
The figure shows the case $n=3$, for a model admitting a good isometry.
In this case, $\Delta_M$ is a triangle with vertices $\eta_1,\eta_2,\eta_3\in 
\partial{\overline {\cal H}}^2=\R\cup \{\infty\}$: these values are displayed on the right. We assume that one reduces to IIA by the good isometry, and that 
$\nu_\epsilon^1=0$ along the edge $[\eta_1,\eta_2]$ and $\nu_3^1=0$ 
for the vertex $\eta_3$ (these conditions are satisfied for the models 
of \cite{Witten_Acharya}, with an appropriate choice of basis for the 
lattice spanned by the toric hyperkahler generators). One obtains a 
three-dimensional cylinder with spherical section 
sitting above the edge $[\eta_1,\eta_2]$ and 
a squashed 
sphere sitting above the vertex $\eta_3$. The two generators of the cylinder 
sitting above the vertices $\eta_1$ and $\eta_2$ are connected with the 
sphere which sits above $\eta_3$ by three-dimensional bodies whose 
generic sections are squashed two-spheres.}
\end{figure}

Above a vertex $\eta_j\neq 0$ (so that $\nu_j^1\neq 0$ and 
$\eta^2\delta\rightarrow \infty$), the induced metric takes the form:
\be
\label{im_segment}
d{\tilde s}^2|_{\theta,\chi,\psi}= \frac{r^2 f}{4}d\theta^2\nn~~,
\ee 
and the fiber reduces to a segment of (relative) length
$\frac{\pi r\sqrt{f}}{2}$. 
If one of the toric hyperkahler generators
has the property $\nu_j^1=0$, then $\eta=0$ is a vertex of $\Delta_M$ above 
which (\ref{im}) reduces to the metric (\ref{im_r}):
\be
\label{im_round}
d{\tilde s}^2|_{\theta,\chi,\psi}= \frac{r^2 f}{4} \left[d\theta^2+
\frac{\sin^2(\theta)}{T^2} d\xi^2\right]\nn~~.
\ee
Above the interior of an edge having $T\neq 0$, the 
fiber has topology $S^2$. The behavior at the edge's endpoints 
depends on whether $\eta$ vanishes there. If $\eta$ does not vanish 
at the endpoint, then the fiber collapses to a segment 
(\ref{im_segment}). If $\eta$ vanishes there, then the 
fiber reduces to the metric (\ref{im_round}). 
Figure \ref{2Ageometry} illustrates this behavior for a particular
example in the case $n=3$. 

\subsubsection{Asymptotics of the coupling constant and RR one-form} 

It is easy to check that the coupling constant has the behavior:
\be
\label{gas}
g_A^{4/3}=\frac{r^2 U_{11}}{2}=\frac{r^2}{4}
\frac{2T^2+f\eta^2\delta \sin^2(\theta)}{L_0^2\delta }+O(\rho)~~.
\ee
For the RR one-form, we obtain:
\bea
\label{Cas}
C_\psi&=&\frac{f\eta \delta\sin^2(\theta)-2TL_1}{
2T^2+f\eta^2\delta\sin^2(\theta)}+O(\rho)~~\nn\\
C_\rho&=&\frac{f\eta L_1\delta\sin\theta\cos\theta\cos\chi}{
2T^2+f\eta^2\delta\sin^2(\theta)}+O(\rho)~~\nn\\
C_\eta&=&~~~~O(\rho)~~\nn\\
C_{u^1}&=&-\frac{f\eta L_0\delta\sin\theta\sin\chi}{
2T^2+f\eta^2\delta\sin^2(\theta)}+O(\rho)~~\nn\\
C_{u^2}&=&\frac{f\eta L_0\delta\sin\theta\cos\chi}{
2T^2+f\eta^2\delta\sin^2(\theta)}+O(\rho)~~.
\eea
Upon converting to spherical coordinates on the $S^2$ fibers spanned 
by ${\vec u}$, one obtains:
\be
\label{Cf}
C=\frac{f\eta \delta\sin^2(\theta)-2TL_1}{
2T^2+f\eta^2\delta\sin^2(\theta)}d\psi+
\frac{f\eta L_1\delta\sin\theta\cos\theta\cos\chi}{
2T^2+f\eta^2\delta\sin^2(\theta)}d\rho+
\frac{f\eta L_0\delta\sin^2(\theta)}{
2T^2+f\eta^2\delta\sin^2(\theta)}d\chi +O(\rho)~~,
\ee
where $\delta = L_0 L_2 - L_1^2$. Changing to the coordinates $\xi,\zeta$
gives:
\be
\label{Cform}
C=\frac{f\eta \delta\sin^2(\theta)-2TL_1}{
2T^2+f\eta^2\delta\sin^2(\theta)}d\xi+
\frac{f\eta L_1\delta\sin\theta\cos\theta\cos\zeta}{
2T^2+f\eta^2\delta\sin^2(\theta)}d\rho+
L_1d\zeta +O(\rho)~~,
\ee
where we used the relation $L_0=\eta L_1+T$. 

Fixing $r,\eta,\theta,\rho$ and $\xi$, one can integrate $C$ over 
the collapsing $\zeta$-circle $\Gamma_\zeta$ with the result: 
\be
\label{Cflux}
\int_{\Gamma_\zeta}{C}=C_\zeta\int_0^{2\pi}{d\zeta}=2\pi L_1 +O(\rho)=
2\pi \nu_\epsilon^2+O(\rho)~~.
\ee
In this relation, we assume that $\rho$ is taken to be very small 
but non-vanishing, so that the cycle $\Gamma_\zeta$ has radius of order $\rho$.
This ensures that the asymptotic form (\ref{Cform}) can be used 
in (\ref{Cflux}). 
Integrating over a similar $\xi$-circle (at $\rho$ small but non-zero)
gives:
\be
\label{Cflux_xi}
\int_{\Gamma_\xi}{C}=C_\xi\int_0^{2\pi}{d\xi}=2\pi
\frac{f\eta \delta\sin^2(\theta)-2TL_1}{
2T^2+f\eta^2\delta\sin^2(\theta)}+O(\rho)~~.
\ee
This shows the existence of flux on the locus $\rho=0$. 

\subsection{Behavior on the horizontal locus} 

We are now ready to study the behavior of the type IIA fields
(\ref{dilaton_C}) and (\ref{metricA}) on the horizontal locus of
Subsection 2.2.  This locus corresponds to $\rho=0$ and $P_\phi= |{\vec
A}_\phi\times {\vec u}|^2=0$
\footnote{As well as $P_{\psi} = 0$ and $P_{\phi \psi} = 0$. These two
equations give the same condition as $P_\phi=0$ in the limit $\rho
\rightarrow 0$, as is clear from (\ref{PQ}).  The vanishing of those
three metric coefficients is an obvious consequence of the fact that
the horizontal locus consists of two-spheres lying in the base $M$
(see Subsection 2.2).}, at some fixed value of $r$. 
By virtue of the asymptotic expansion of 
${\vec A}_\phi$ given in (\ref{Aas}) and the constraint $|{\vec
u}|=1$, the second condition amounts to 
${\vec u}=(0,0,1)$, which due to (\ref{sphcoord}) is equivalent 
with $\theta=0,\pi$. These two values of $\theta$ correspond to the two
opposite edges of 
the characteristic polygon $\Delta$ lying above a given edge of $\Delta_M$
(see Section 3).
Along this locus, the quantities $U_{ij}$ coincide with $B,D,E$.

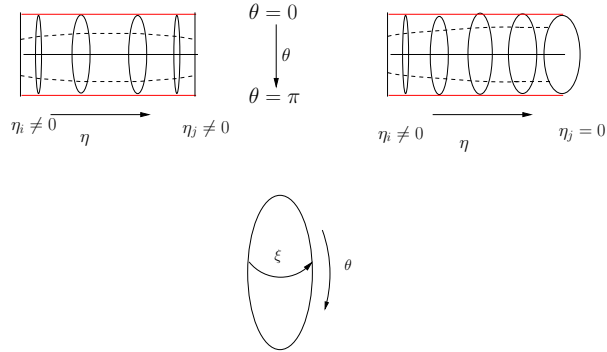
\begin{figure}[hbtp]
\begin{center}
\scalebox{0.4}{\input{hor_strong_a.pstex_t}}
\end{center}
\caption{\label{hor_strong_a} The locus $\theta=0,\pi$ at $\rho=0$ and 
fixed $r$ for an edge of $\Delta_M$ which satisfies 
$\nu_\epsilon^1\neq 0$. Since ${\tilde R}_\zeta=0$ above this edge and 
since ${\tilde R}_\xi$ vanishes at $\theta=0,\pi$, we 
obtain two segments which are 
strongly coupled loci in IIA string theory. The $(\theta,\xi)$-fiber 
over the given edge is a squashed two-sphere, which reduces to a vertical 
segment above the edge's endpoints, unless one of 
the endpoints lies at $\eta=0$, in which case the fiber does not degenerate
there. These two cases are shown in the upper figures.
The horizontal segments result 
by fibering the north and south poles of the $(\theta,\xi)$-fiber 
over the edge.  
The lower figure shows the coordinates 
along a generic $(\theta,\xi)$ fiber.}\end{figure}

From expression (\ref{detUas}), one finds
\footnote{Indeed, the terms of order $\rho^0$ 
and $\rho$ vanish while $\Lambda$ reduces to $2(2\delta+fL_1^2)$ .}: 
\be
\label{detU0}
det(U)=\frac{1}{L_0^4}\left[1+\frac{fL_1^2}{2\delta}\right]
\rho^2+O(\rho^3)~~{\rm ~for~}\theta=0,\pi~~.
\ee
Equation (\ref{gas}) gives the behavior
of the coupling constant:
\be
\label{gas0}
g_A^{4/3}=\frac{r^2}{2}\frac{T^2}{L_0^2\delta}+O(\rho)~~{\rm for}
~~\theta=0,\pi~~.
\ee
Let us fix an edge $c$ of $\Delta_M$, and thereby (as explained in Section 3) 
the value of the sign vector $\epsilon$. We let $e$ and $-e$ be the two edges 
of $\Delta$ lying above $c$, which correspond respectively to the loci 
$\theta=0$ and $\theta=\pi$ at $\rho=0$ and fixed $r$.
We obtain the following two cases:

\paragraph{(a)} 
If the edge $c$ satisfies $T=\nu_{\epsilon(\eta)}^1\neq 0$, 
then equation (\ref{im}) shows
that the $(\theta,\zeta)$-
fibers above $c$ are squashed two-spheres (figure \ref{hor_strong_a}). 
Further imposing the conditions $\theta=0,\pi$ gives two segments 
lying above the edge $c$. Relation (\ref{Cform}) gives:
\be
C=-\frac{L_1}{T}d\xi+L_1d\zeta+O(\rho)~~{\rm~for~}\theta=0,\pi~~,
\ee
while (\ref{Cflux}) becomes:
\be
\int_{\Gamma_\zeta}{C}=2\pi \nu_\epsilon^2 +O(\rho)~~{\rm~for~} \theta=0,\pi~~.
\ee
This shows the existence of $L_1=\nu_\epsilon^2$ units of flux passing 
through the poles of each $\theta,\xi$-sphere lying at $\rho=0$.
Equation (\ref{gas0}) shows that the coupling constant tends to a finite 
value on this piece of the horizontal locus, which therefore 
does not admit a perturbative interpretation in IIA string theory.

\noindent

\begin{figure}[hbtp]
\begin{center}
\scalebox{0.5}{\input{hlocus.pstex_t}}
\end{center}
\caption{\label{hlocus} Schematic depiction of the horizontal branes sitting 
above the edge $c$ of $\Delta_M$ for the case $\nu_\epsilon^1=0$. 
The locus $\rho=0$ at a fixed value of $r$ is a 3-dimensional 
cylinder whose cross section is a two-sphere spanned by $\eta$ and $\xi$. 
The internal directions of each D6-brane 
comprise $r$ and one of the horizontal 
two-spheres which bound this cylinder. The figure also shows a cross section 
of the three-dimensional 
cylinder at a fixed value of $\eta$. This cross section is a two-dimensional 
cylinder with coordinates $\theta$ and $\xi$, the degeneration of the 
$(\theta,\xi)$-fiber above the locus $\rho=0$.}
\end{figure}
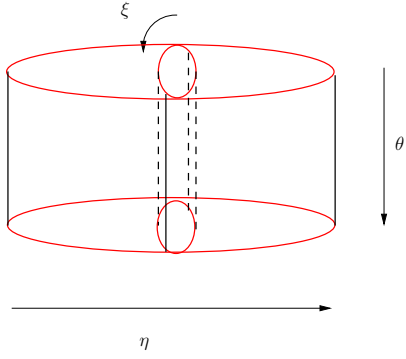

\paragraph{(b)} 
If $T=\nu_{\epsilon}^1$ vanishes, then (\ref{ABDE_as}) gives $B=E=0$
for $\rho\rightarrow 0$. 
Equation (\ref{im_cyl}) shows that the $(\theta,\xi)$ fibers degenerate 
to two-dimensional cylinders in this limit.
As explained above, the part of the locus $\rho=0$ which sits above 
the edge $c$ is a three-dimensional 
cylinder of finite height whose cross section 
is a two-sphere spanned by $\eta$ and $\xi$.
The sublocus $\theta=0,\pi$ consists of the two spherical ends 
of this cylinder (figure \ref{hlocus}).

\noindent

Relation (\ref{gas}) shows that the 
type IIA coupling constant vanishes on these ends: 
\be
g_A^{4/3}=\frac{r^2f\eta^2}{4L_0^2}\sin^2(\theta)+O(\rho)~~.
\ee
Since $T=0$, equation (\ref{Cform}) reduces to:
\be
\label{Cform0}
C=\frac{1}{\eta}d\xi+\frac{L_1\cos\zeta}{
\eta\tan\theta} d\rho+L_1d\zeta+O(\rho)~~,
\ee
so that $C_\rho$ blows up for $\rho=0$ and 
$\theta=0,\pi$ while $C_\xi$ and $C_\zeta$ remain 
finite. Relation (\ref{Cflux}) gives:
\be
\label{Cflux0}
\int_{\Gamma_\zeta}{C}=2\pi L_1=2\pi \nu_\epsilon^2~~.
\ee
In this relation, we integrate along the circle $\Gamma_\zeta$ 
obtained by fixing $\theta=0$ or $\pi $ as 
well as $r, \xi$ and $\eta$, and a small 
(but non-vanishing) value of $\rho$.
Equation (\ref{Cflux0}) indicates the presence of RR flux 
through the 2-dimensional disk $D_\zeta$ bounded by $\Gamma_\zeta$: 
\be
\label{flux0}
\int_{D_\zeta}{F_2}=2\pi \nu_\epsilon^2~~,
\ee
where $F_2:=dC$. Since the radius of such a $\zeta$-circle can be 
made arbitrarily small, this equation indicates the presence of a source. 
The flux sources 
must be localized at $\theta=0$ and $\theta=\pi$ since $C_\rho$ becomes 
singular there. 

This establishes the presence of  two $D6$-branes 
(of multiplicity $|\nu_\epsilon^2|$) at $\theta=0,\pi$. 
Their internal directions span $r$, $\eta$ and $\xi$, which for 
$r_0=0$ are the cones 
over the two-spheres bounding the cylinder in figure \ref{hlocus}.
In this case, the full worldvolume 
of each such brane is topologically 
a copy of $\R^7$ spanned by the external ${\bf E}^{3,1}$ 
coordinates and the internal directions mentioned above.
Vanishing of $T=\nu_\epsilon^1$ 
implies that $|\nu_\epsilon^2|=gcd(0,\nu_\epsilon^2)=
gcd(\nu_\epsilon^1,\nu_\epsilon^2)$ coincides with the integer 
$m_e=m_{-e}$ introduced in \cite{toric} and reviewed in Subsection 2.2
($e$ is one of the two edges of $\Delta$ lying above the given edge 
of $\Delta_M$). Therefore, 
the D6-branes sitting at $\theta=0$ and $\theta=\pi$ 
have multiplicity $|m_e|$, and carry an $SU(|m_e|)$ gauge theory 
on their worldvolume. This agrees 
with our expectations and with the conclusions of \cite{toric}. 
Since we divided by the isometry associated with $\phi$, we find 
D6-branes corresponding 
to those singular loci of $Y_H$ which are fixed by this isometry  -- these 
are the horizontal loci for which $T=\nu_\epsilon^1=0$. 

\subsection{Behavior on the vertical locus}

Recall from Section 3 that on the vertical locus one has $\rho=0$ 
and $\eta=\eta_j:=
-\frac{\nu^1_j}{\nu^2_j}$ 
for some $j$. Thus $L_2=\frac{|\nu_j^2|}{|\eta-\eta_j|}+O(\eta-\eta_j)$
tends to infinity on this locus. 
Then equations (\ref{Uas}) give:
\be
\label{Uas1}
U_{11}=\frac{f\eta_j^2}{2L_0^2}\sin^2(\theta)~~,~~ 
U_{22}=\frac{f}{2L_0^2}\sin^2(\theta)~~,
~~U_{12}=\frac{f\eta_j}{2L_0^2}\sin^2(\theta)~~,
\ee
while the determinant of $U$ vanishes. Equation (\ref{gas}) gives:
\be
g_A^{4/3}=\frac{r^2f\eta_j^2}{4L_0^2}\sin^2(\theta) + O(\rho)~~.
\ee 

The behavior of $C$ results by taking the limit $\eta\rightarrow \eta_j$ in 
equations (\ref{Cform}). Let us fix some  $j=1\dots n$ and concentrate 
on the piece of the vertical locus lying above the vertex 
$P_j$ of $\Delta_M$ given by $\rho=0$ and $\eta=\eta_j$.
Once again, we have two possibilities:

\paragraph{(a)} 
$\eta_j\neq 0\Leftrightarrow \nu_j^1\neq 0$. In this case  
$g_A$ has a non-vanishing  limit for $\rho\rightarrow 0$
and the loci $\theta=0,\pi$ are strongly coupled. 
Equation (\ref{im_segment}) shows that that the $(\theta,\xi)$-fiber degenerates
to a segment of (relative) length $\frac{\pi r}{2}\sqrt{f}$.

Equation (\ref{Cform}) gives:
\be
C=\frac{1}{\eta_j}d\xi+\frac{L_1\cos\zeta}{
\eta_j\tan\theta} d\rho+L_1d\zeta+O(\rho)~~.
\ee

\paragraph{(b)}
 $\eta_j=0\Leftrightarrow \nu_j^1=0$. In this case the coupling constant 
$g_A$ vanishes on the fiber $S^2_j(\theta,\xi)$ sitting above $\rho=\eta=0$, 
which carries the metric (\ref{im_round}).
The behavior of the RR one-form is more subtle than before, because
(\ref{Cform}) contains the combination $\eta L_2$. To find its limit for 
$\eta \rightarrow 0$, notice that $\epsilon_k(0)=sign(\nu_k^1)$ for $k\neq j$ 
and $\epsilon_j(\eta)=sign(\eta)sign(\nu_j^2)$. The limit of 
the latter depends on whether $\eta$ approaches zero 
from below or from above. If we let $\kappa=sign(\eta)$, then the two 
directional limits correspond to $\kappa=+1$ and $\kappa=-1$ respectively. 
Since $\nu_j^1$ vanishes, we have $L_0(0)=\sum_{k\neq j}{|\nu_k^1|}$ and  
$T(0)=\sum_{k\neq j}{\epsilon_k(0)\nu_k^1}=\sum_{k\neq j}{
sign(\nu_k^1)\nu_k^1}=\sum_{k\neq j}{|\nu_k^1|}=L_0(0)$. Upon defining the 
vector ${\tilde \nu}_j:=\sum_{k\neq j}{\epsilon_k \nu_k}=
\sum_{k\neq j}{sign(\nu_k^1)\nu_k}$, we obtain:
\be
\label{L0T0}
L_0(0)=T(0)={\tilde \nu}_j^1~~. \label{T0}
\ee
Notice that $T(\eta)$ equals $T(0)$ on an entire vicinity of $\eta=0$ 
(this follows using the fact that $T(0)$ is a locally constant function). 
On the other hand, we have $L_1=\epsilon_j\nu_j^2+
\sum_{k\neq j}{\epsilon_k\nu_k^2}$, so that:
\be
\lim_{\tiny \begin{array}{c}\eta\rightarrow 0\\sign(\eta)=\kappa\end{array}} 
{L_1(\eta)}=\kappa |\nu_j^2|+{\tilde \nu}_j^2~~.
\ee
The quantity 
$L_2(\eta)=\sum_{k=1}^n{\frac{(\nu_k^2)^2}{|\nu_k^2\eta+\nu_k^1|}}$
has the following behavior for $\eta\rightarrow 0$:
\be
L_2(\eta)=\kappa \frac{|\nu_j^2|}{\eta}+\sum_{k\neq j}{
\frac{(\nu_k^2)^2}{|\nu_k^1|}}
+O(\eta)~~.
\ee
This implies:
\bea
\lim_{\tiny \begin{array}{c}\eta\rightarrow 0\\sign(\eta)=\kappa\end{array}}
{\eta L_2}=\kappa |\nu_j^2|~~. \label{eL}
\eea

Using (\ref{Cform}), we obtain: 
\bea
C_\xi^{(\kappa)}&:=& 
\lim_{\tiny \begin{array}{c}\eta\rightarrow 0\\sign(\eta)=\kappa\end{array}}
{C_\xi} = -\kappa |\nu_j^2|
\frac{1-\frac{f}{2}\sin^2(\theta)}{{\tilde \nu}_j^1}-\frac{\tilde{\nu}_j^2}{\tilde{\nu}_j^1}~~\nn\\
C_\rho^{(\kappa)}&:=& \lim_{\tiny \begin{array}{c}\eta\rightarrow 0\\sign(\eta)=\kappa\end{array}}
{C_\rho}=\left(\frac{f}{2}\frac{(\nu_j^2)^2}{{\tilde \nu}_j^1}+
\kappa |\nu_j^2|\frac{f}{2}\frac{{\tilde \nu}^2}{{\tilde \nu}^1}\right)\sin(\theta)\cos(\theta)\cos(\zeta)~~\\
C_\zeta^{(\kappa)}&:=& 
\lim_{\tiny \begin{array}{c}\eta\rightarrow 0\\sign(\eta)=\kappa\end{array}} 
{C_\zeta}={\tilde \nu}_j^2+\kappa |\nu_j^2|~~.\nn
\eea
These expressions show that $C$ 
is discontinuous along the vertical locus $\eta=\eta_j=0$. The 
discontinuity jumps of $C_\xi$ and $C_\zeta$ are given by:
\bea
\label{Clim_zeta}
(\Delta C)_\xi:&=&
\lim_{\tiny \begin{array}{c}\eta\rightarrow 0\\ \eta>0\end{array}}C_\xi-
\lim_{\tiny \begin{array}{c}\eta\rightarrow 0\\ \eta<0\end{array}}C_\xi=
-2\frac{|\nu_j^2|}{
{\tilde \nu}_j^1}\left[1-\frac{f}{2}\sin^2(\theta)\right]~~, \nn \\
(\Delta C)_\zeta:&=&
\lim_{\tiny \begin{array}{c}\eta\rightarrow 0\\ \eta>0\end{array}}C_\zeta-
\lim_{\tiny \begin{array}{c}\eta\rightarrow 0\\ \eta<0\end{array}}C_\zeta=
2|\nu_j^2| ~~.
\eea
Since the IIA coupling vanishes on $S^2_j(\theta,\xi)$, 
we expect a D6-brane 
whose worldvolume comprises ${\bf E}^{3,1}$ and the cone over
$S^2_j$. To confirm this picture, we shall 
compute the `magnetic' RR flux produced by 
our brane. 

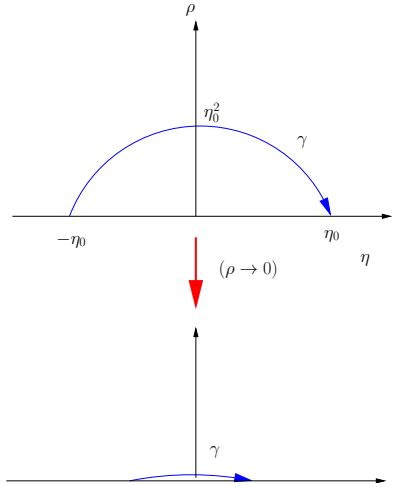
\begin{figure}[hbtp]
\begin{center}
\scalebox{0.4}{\input{curve.pstex_t}}
\end{center}
\caption{\label{curve} The curve $\gamma$. The parameter $\eta_0$ is 
taken to be of order $\rho$ (for example, $\eta_0=\rho$). 
As $\rho$ tends to zero, the point $\eta_0^2$ approaches zero faster than 
$\eta_0$. As a consequence, the curve $\gamma$ 
looks like an infinitesimally 
small segment along the $\eta$-axis in this limit. }
\end{figure}

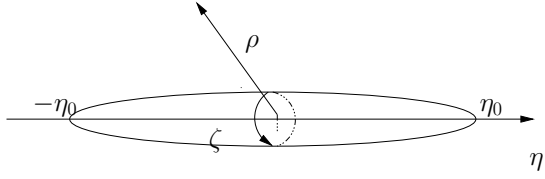
\begin{figure}[hbtp]
\begin{center}
\scalebox{0.6}{\input{surface_a.pstex_t}}
\end{center}
\caption{\label{surface_a} The two-sphere 
$\Sigma$ determined by $\gamma$ in the 
directions transverse to the worldvolume. The internal 
worldvolume coordinates $r, \theta$ and 
$\xi$ are fixed to some arbitrary values. As $\rho$ tends to zero, the 
$\zeta$-circle collapses faster than $\eta_0$. The point $(r,\theta,\xi)$ 
of the internal worldvolume sits at $\eta_0=\rho=0$. }
\end{figure}

The charge computation requires us to fix the internal worldvolume 
coordinates $r,\theta, \xi$ to some values 
and find a closed surface sitting in the transverse directions 
$\rho,\eta,\zeta$ and surrounding the worldvolume 
point $(r,\theta,\xi)$. For this, consider the curve
$\gamma$ given by the following equation in coordinates $(\rho,\eta)$:
\be
\rho=\eta_0^2-\eta^2~~.
\ee
Here $\eta_0$ is a fixed parameter. Since we wish to use the asymptotic 
expressions determined above (which are valid only to leading order in 
$\rho$), we will take $\eta_0$ to be very small, so that the curve sits in a 
small vicinity of $\rho=\eta=0$ (figure \ref{curve} ). Inclusion of the 
remaining transverse coordinate (namely $\zeta$) gives an $S^1$ fibration 
over this curve, whose fiber collapses to a point above the endpoints 
$\rho=0,\eta=\pm \eta_0$. This gives a surface 
$\Sigma$ (topologically a two-sphere) which surrounds the worldvolume point 
$(r,\theta,\xi)$ sitting at $\rho=\eta=0$ (figure \ref{surface_a}).

\noindent

\noindent

The magnetic flux of $F_2=dC$ through $\Sigma$ is given by the integral
$\Phi=\int_{\Sigma}{F_2}$. To compute this, one should in 
principle start with relation (\ref{Cform}) 
and perform the integral of $dC$ over $\Sigma$, 
while keeping in mind that $\eta_0$
is taken to be very small. It is easy to find the result without 
performing the entire computation. For this, notice that the 
integral only receives
contributions from the components $F_{\zeta\eta}$ and $F_{\zeta\rho}$ 
(the second is relevant because $d\rho$ and $d\eta$ are related on the surface 
$\Sigma$, which is also why $F_{\rho \, \eta}$ gives zero contribution). Using the defining equation of $\gamma$, one finds 
$d\rho=-2\eta d\eta$ on $\Sigma$, which means that, after pull-back 
to this surface, the component $F_{\zeta\rho}$ will be multiplied by $\eta$. 
Since $\eta$ runs between $-\eta_0$ and $\eta_0$ and 
since 
$\eta_0$ is taken to zero in the end, this component does not contribute to 
the final result. Therefore, it suffices to consider the contribution 
coming from $F_{\zeta\eta}$, which gives:
\be
\Phi=\int_{\Sigma}{F_{\zeta\eta}d\zeta\wedge d\eta}=
\int{d\zeta{\int_{-\eta_0}^{\eta_0}{d\eta}
\frac{\partial C_\zeta}{\partial \eta}}}=\int{d\zeta [C_\zeta(\eta_0)-
C_\zeta(-\eta_0)]}~~.
\ee
Taking the limit $\eta_0\rightarrow 0$, the integrand reduces to 
$(\Delta C)_\zeta=C_\zeta^{(+)}-C_\zeta^{(-)}$, so that:
\be
\label{Phi_0}
\Phi=\int_{0}^{2\pi}{d\zeta (\Delta C)_\zeta}~~.
\ee
Using the last of equations (\ref{Clim_zeta}), we find:
\be
\label{Cflux_vert}
\Phi=4\pi |\nu_j^2|~~.
\ee
As for the horizontal locus, we find 
agreement with our expectations and with the results of
\cite{toric}: one obtains localized D6-branes for a vertical locus which
satisfies $\nu_j^1=0$ and 
thus is fixed under the action of $U(1)_{\phi}$.

\paragraph{Observation} If one takes into account the possibility of  
further $\Z_2$ identifications above certain edges or vertices 
of $\Delta_M$ (see Observation 3 in Subsection 3), then 
the reduction to IIA will produce an 
identification $\zeta\equiv \zeta+\pi$ on such a locus. In this case, the 
range of integration in relations (\ref{Cflux0}) and (\ref{Phi_0}) 
must be  restricted 
to $[0,\pi]$, which means that the result in the right hand side of these 
relations must be divided by two. This agrees with the results of 
\cite{toric}, which show that the gauge group content may be half 
of that predicted by (\ref{Cflux0}) and (\ref{Phi_0}) 
in certain situations. 
Precise criteria for deciding when this happens 
can be found in \cite{toric}.

\section{T-dual IIB description}
\setcounter{equation}{0}

\subsection{The type IIB solution} 

The metric (\ref{IIA}) is still invariant under shifts of the
coordinate $\psi$. In this section we will T-dualize along
the direction of this coordinate. Applying the Buscher rules
\cite{Buscher}, \cite{BHO} we obtain the following IIB fields in the
string frame for the NS-NS sector: 
\bea
\label{phiB}
\varphi_{B} &=& \ln U_{11} - \frac{1}{2} \ln \det U \nn\\ 
B^{NS} &=&
g_{\psi m}^{(A)} \, g_{\psi \psi}^{(A)-1} \, d \psi \wedge dx^m  \\
ds_{(B)}^2 &=& [g_{mn}^{(A)} - g_{\psi \psi}^{(A)-1} \, g_{\psi m}^{(A)} \, g_{\psi
n}^{(A)}] \, dx^m dx^n + g_{\psi \psi}^{(A)-1} \, d \psi^2 ~~.\nn 
\eea
Here $x^m$ denote the 9 coordinates different from $\psi$.  
Since the IIA background contains no
$B$-field, 
the T-duality rules mapping it to IIB are simpler than the general ones.

In the R-R sector, we obtain:
\bea
l~ &=& C_{\psi} = \frac{U_{12}}{U_{11}} \nn \\
B^R &=& [C_m - g_{\psi \psi}^{(A)-1} \, g_{\psi m}^{(A)} \, C_{\psi}] 
dx^m \wedge d \psi, \label{RRb}
\eea
where $l$ is the axion and $B^R$ the RR 2-form. The IIB background does 
not contain a RR 3-form, because of the vanishing $B$-field in the 
IIA solution.

We remind the reader that the relation between string and Einstein frames is:
\be
ds_E^2 = e^{-\frac{1}{2} \varphi} \, ds_{S}^2 \, .
\ee
As in the previous section, it will be convenient to use the 
`relative' type IIB metric ${\tilde g}_{ij}^{(B)}$, related to $g_{ij}^{(B)}$
through:
\be
g_{ij}^{(B)}=\left(\frac{r^2U_{11}}{2}\right)^{1/2}{\tilde g}_{ij}^{(B)}~~.
\ee

The IIB coupling results from relation (\ref{phiB}):
\be
g_B=e^{\varphi_B}=\frac{U_{11}}{\sqrt{det U}}~~.
\ee
Hence the axion-dilaton modulus is $\tau=l+\frac{i}{g_B}=
\frac{U_{12}+i\sqrt{det U}}{U_{11}}$. 
Recall that IIB string theory has a 
non-perturbative duality symmetry which acts on 
$\tau$ and on the two-form fields through:
\be
\label{SL2Z}
\tau\rightarrow \frac{a\tau+b}{c\tau +d}~~,~~
{\cal B} \rightarrow 
S^{-T}{\cal B}~~,
\ee
where $S=\left[\begin{array}{cc}a&b\\c&d\end{array}\right]$ is an 
$SL(2,\Z)$ matrix and 
${\cal B}=\left[\begin{array}{c} B^{NS}\\B^R\end{array}\right]$.

\subsection{Behavior of the IIB solution for $\rho\rightarrow 0$}
\setcounter{equation}{0}

We are now ready to study the
behavior of the type IIB solution
near the positions of the horizontal and vertical loci. 
We once again start by extracting the asymptotics of various
fields in the limit $\rho\rightarrow 0$.

\subsubsection{Asymptotics for the metric induced in the 
$\theta,\chi,\psi$ directions}

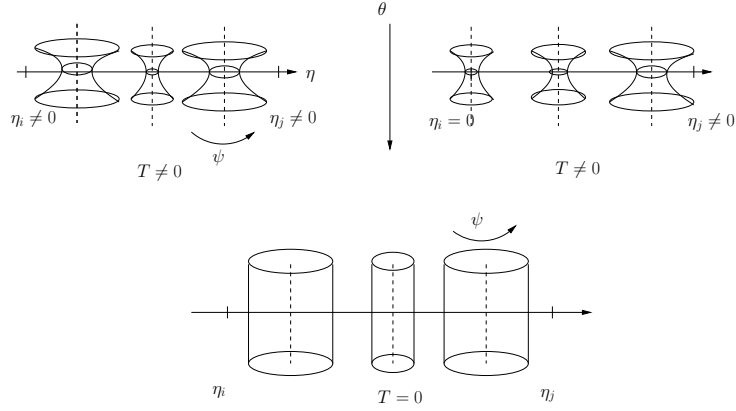
\begin{figure}[hbtp]
\begin{center}
\scalebox{0.4}{\input{b_degeneration.pstex_t}}
\end{center}
\caption{\label{b_degeneration} 
The IIB $(\theta,\psi)$-fibers above various edges of $\Delta_M$.}
\end{figure}

Using relations (\ref{phiB}) and (\ref{angularas}), we compute:
\bea
\label{RpsiB}
{\tilde g}^{(B)}_{\psi\psi}&=&\frac{2\eta^2 \delta }{r^2}+\frac{4}{r^2 f}\frac{T^2} {\sin^2(\theta)} + O(\rho) \nn\\ 
{\tilde g}^{(B)}_{\theta \theta}&=&{\tilde g}^{(A)}_{\theta
\theta}=\frac{r^2 f}{4} + O(\rho^2) \nn \\ {\tilde g}^{(B)}_{\theta
\chi} &=& O(\rho) \nn\\ 
{\tilde g}^{(B)}_{\chi \chi} &=& O(\rho)~~,\nn\\
{\tilde g}^{(B)}_{\psi\chi}&=&{\tilde g}^{(B)}_{\psi\theta}=0~~.
\eea
Note that the $\chi$-circle is always collapsed to a point for $\rho=0$.
We have:
\be
\label{sqB}
d{\tilde s}^2_B|_{\theta,\chi,\psi}=
\frac{r^2 f}{4} d\theta^2+\frac{2}{r^2f}
\frac{2T^2+f\eta^2\delta\sin^2(\theta)}{\sin^2(\theta)} 
d\psi^2 + O(\rho)~~,
\ee 
which shows that the locus $\rho=0$ has codimension two. 
In particular, the $\theta$-fiber 
degenerates to a segment of (relative) length 
$\frac{\pi r\sqrt{f}}{2}$ 
above the entire boundary of 
the hyperbolic plane.
The $\psi$-circle fibers over this segment with the
radius given by (\ref{sqB}), while the $\chi$-circle is collapsed 
to zero size. Above an edge having $T=0$, 
(\ref{sqB}) reduces to:
\be
\label{sqB0}
d{\tilde s}^2_B|_{\theta,\chi,\psi}=
\frac{r^2 f}{4} d\theta^2+
\frac{2\eta^2\delta}{r^2} 
d\psi^2 + O(\rho)~~,
\ee
which shows that the $(\theta,\chi,\psi)$-fibers are cylinders of 
squared relative radius $\frac{2\eta^2\delta}{r^2}$; this radius tends to 
infinity at the two endpoints of the given edge of $\Delta_M$ (none of which 
can correspond to $\eta=0$).  Above an edge with $T\neq 0$, the 
squared $\psi$-circle
radius varies with $\theta$ and is symmetric under 
$\theta\rightarrow \pi -\theta$. For fixed $\eta$, 
it tends to an infinite value for 
$\theta=0,\pi$ and reaches its minimum $\frac{2(f\eta^2\delta+2T^2)}{r^2f}$ 
for $\theta=\pi/2$. This squared radius tends to infinity for any $\theta$ 
at a vertex $\eta_j\neq 0$, and 
reduces to $\frac{4T^2}{r^2f\sin^2(\theta)}$ for $\eta=0$. The situation is 
summarized in figure \ref{b_degeneration}. 

\noindent

\subsubsection{Asymptotic behavior of the modular parameter and NS-NS/RR
two-forms}

The modular parameter has the following asymptotics for $\rho\rightarrow 0$:
\bea
\label{tau_as}
Re\tau&=&\frac{-2L_1T+f\eta\delta\sin^2(\theta)}{2T^2+f\eta^2\delta
\sin^2(\theta)}+O(\rho)\nn\\
Im\tau&=&\frac{L_0\sqrt{2f\delta}\sin(\theta)}{2T^2+f\eta^2\delta
\sin^2(\theta)}+O(\rho)~~.  \eea
For the NS-NS and R-R two-forms, one obtains:
\bea
\label{Bas}
{\cal B}_{\psi \rho}&=&\nu_\epsilon \frac{L_1}{L_0}
\frac{\cos \chi}{\tan \theta}+O(\rho)~~\nn\\
{\cal B}_{\psi\eta}&=&O(\rho)\nn\\
{\cal B}_{\psi u^1}&=&-\nu_\epsilon \frac{\sin \chi}{\sin \theta}+O(\rho)~~\\
{\cal B}_{\psi u^2}&=&\nu_\epsilon\frac{\cos \chi}{\sin \theta}+O(\rho)~~\nn~~,
\eea
where we used $T=\nu_\epsilon^1$ and $L_1=\nu_\epsilon^2$.
Converting to the coordinates $(\rho,\eta,\psi,\theta,\chi)$, we obtain:
\be
\label{Bform}
{\cal B}=
\nu_\epsilon \left[\frac{L_1}{L_0}\frac{\cos \chi}{\tan \theta} 
d\psi\wedge d\rho+d\psi \wedge d\chi\right]+O(\rho)~~.
\ee
Integration over a $(\chi,\psi)$-torus $\Gamma_{\chi,\psi}$ at a small 
but non-vanishing $\rho$  gives $\nu_\epsilon^1$ units of 
NS-NS flux and $\nu_\epsilon^2$ units of R-R flux through the 
3-dimensional volume $S_{\chi,\psi}$ with boundary $\Gamma_{\chi,\psi}$:
\be
\label{Bflux}
\int_{S_{\chi,\psi}}{d{\cal B}}=\int_{\Gamma_{\chi,\psi}}
{{\cal B}}=(2\pi)^2 \nu_\epsilon~~.
\ee

As in \cite{toric}, we write $\nu_\epsilon=m_\epsilon t_\epsilon$, where 
$m_\epsilon=gcd(\nu_\epsilon^1,\nu_\epsilon^2)$ and 
$t_\epsilon$ is a primitive integral vector. Since $t_\epsilon^1$ and 
$t_\epsilon^2$ are
coprime integers, there exist coprime integers $a$ and $b$ such that 
$a t_\epsilon^2-bt_\epsilon^1=1$. Then the matrix 
$S=\left[\begin{array}{cc}a&b\\t_\epsilon^1&t_\epsilon^2\end{array}\right]$ belongs to 
$SL(2,\Z)$ and has the property:
\be
S^{-T}\nu_\epsilon=\left[\begin{array}{c}0\\m_\epsilon\end{array}\right]~~.
\ee
Using this transformation in (\ref{SL2Z}) allows us to bring $B^{NS}$ and $B^R$ 
to the forms:
\be
B^{'NS}=O(\rho)
\ee
and
\bea
\label{Bprime}
B^{'R}=m_\epsilon \left[\frac{L_1}{L_0}\frac{\cos \chi}{\tan \theta} 
d\psi\wedge d\rho+d\psi \wedge d\chi\right]+O(\rho)~~.
\eea
After this transformation, the modular parameter (\ref{tau_as}) becomes:
\be
\tau'=\frac{a \, [-2L_1 T + (f\eta \delta + i L_0 \sqrt{2f \delta}) \sin^2 (\theta)]+b \, [2T^2 + f \eta^2 \delta \sin^2 (\theta)]}{t_{\epsilon}^1 \, [-2L_1 T + (f\eta \delta + i L_0 \sqrt{2f \delta}) \sin^2 (\theta)] + t_{\epsilon}^2 \, [2T^2 + f \eta^2 \delta \sin^2 (\theta)]} + O(\rho)~~.
\ee

\subsection{Behavior on the horizontal locus}

Recall that the horizontal locus is obtained for $\rho=0$ 
and $\theta=0,\pi$.
Fixing an edge of $\Delta_M$, we distinguish two situations:

\paragraph{(a) $T\neq 0\Leftrightarrow \nu_\epsilon^1\neq 0$.}

In this case, the radius of the $\psi$-circle blows up at $\theta=0,\pi$
and reaches its minimum for $\theta=\pi/2$.
Equation (\ref{tau_as}) gives: 
\be
\label{tau1}
Re(\tau)=-\frac{L_1}{T}+O(\rho)=-\frac{\nu_\epsilon^2}{\nu_\epsilon^1}
+O(\rho)~~,~~Im(\tau)=O(\rho)~~{\rm ~for~}~~
\theta=0,\pi~~.
\ee
In particular, the IIB coupling constant $g_B$ tends to infinity, while 
the axion reaches a finite value. Equation (\ref{Bflux}) shows
\footnote{One may wonder whether the flux computation of 
(\ref{Bflux}) can be applied to the current situation in which $\psi$
is a worldvolume coordinate. Since a delocalized 5-brane can be viewed
as $N$ localized 5-branes at positions $\psi_0, \psi_0 + \frac{2
\pi}{N}, ..., \psi_0 + (N-1) \frac{2 \pi}{N}$ along the $\psi$-circle
in the limit $N \rightarrow \infty$, in the calculation of the 5-brane
flux we can view the $\psi$ direction as transverse rather than
longitudinal direction. We will use this point of view 
again in the remainder of this section.}
the presence 
of NS5 and D5-brane charge on this locus. Due to symmetry of the 
asymptotic solution with respect to shifts of $\psi$, the worldvolume 
directions are spanned by 
${\bf E}^{3,1}$ together with $r,\eta$ and $\psi$,
which means that the 5-branes are {\em delocalized} in the direction $\psi$.

\paragraph{(b) $T=0\Leftrightarrow \nu_\epsilon^1=0$.}

In this case (\ref{RpsiB}) gives:
\be
{\tilde g}_{\psi\psi}^{(B)}=\frac{2\eta^2}{r^2}\delta+O(\rho)~~,
\ee
so that the (relative) radius of the $\psi$-circle remains finite. 
Relations (\ref{tau_as}) give:
\be
Re\tau=\frac{1}{\eta}+O(\rho)~~,~~Im\tau=\frac{L_0}{\eta^2\sin\theta}
\sqrt{\frac{2}{f \delta }}+O(\rho)~~.
\ee
In particular, the IIB coupling constant vanishes for $\rho=0$ 
and $\theta=0,\pi$, so that the horizontal locus is weakly coupled. 
Since $T=0$, one has $L_0=\eta L_1$, and equation (\ref{Bform}) becomes:
\be
{\cal B}=\nu_\epsilon \left[\frac{1}{\eta}\frac{\cos \chi}{\tan \theta} 
d\psi\wedge d\rho+
d\psi \wedge d\chi\right]+O(\rho)~~.
\ee
Equation (\ref{Bflux}) gives:
\be
\label{Phi_Bh}
\int_{S_{\chi,\psi}}{dB^{NS}}=0~~,~~\int_{S_{\chi,\psi}}{dB^{R}}=(2\pi)^2 
\nu_\epsilon^2~~.
\ee
This shows that the horizontal locus contains a $D5$-brane of multiplicity 
$|\nu_\epsilon^2|=m_\epsilon=gcd(0,\nu_\epsilon^2)$. 
The worldvolume directions are ${\bf E}^{3,1}$ and 
$r,\eta, \psi$, which shows that 
the D5-brane is {\em delocalized along $\psi$}. Since the T-dual 
horizontal D6-brane is localized for $T=0$, this delocalization is 
standard: one has performed a T-duality along the direction $\psi$ which 
lies along the worldvolume of the IIA D6-brane
\footnote{From the worldsheet 
perspective, delocalization appears due to the fact that the von Neumann 
conditions $\partial_n X^i=0$ (where $X^i$ are the transverse coordinates) 
are purely differential, while the Dirichlet 
conditions contain a differential part $\partial_\tau X^i=0$ and an `integral'
part $X^i_{CM}=a^i$ (where $X_{CM}^i$ are the string 
center of mass coordinates and 
$a^i$ characterize the position of the D-brane in the transverse directions). 
In general, T-duality translates the von Neumann conditions 
into the differential part of the Dirichlet conditions, which fixes the 
boundary behavior of the oscillator modes but not that of the center of mass mode. 
Except for particular backgrounds, the condition $X_{CM}^i=a^i$ cannot be 
recovered by this process, which means that the dual 
D-brane is delocalized (i.e. $a^i$ can take arbitrary values).}. 

\subsection{Behavior on the vertical locus}
Let us consider  the vertical locus 
$\rho=0,\eta\rightarrow 
\eta_j:=-\frac{\nu_j^1}{\nu_j^2}$. Once again, we distinguish two situations:

\paragraph{(a) $\eta_j\neq 0\Leftrightarrow \nu_j^1\neq 0$.}

In this case, equation (\ref{RpsiB}) shows that the (relative) radius of the 
$\psi$ circle tends to infinity for $\eta\rightarrow \eta_j$. Relation
(\ref{tau_as}) gives: 
\be
Re\tau=\frac{1}{\eta_j}=-\frac{\nu_j^2}{\nu_j^1}~~,~~Im\tau=0~~.
\ee
The second equation shows that the 
IIB coupling constant tends to infinity. Therefore, this locus 
does not admit a perturbative interpretation in IIB string theory. 

\paragraph{(b) $\eta_j=0\Leftrightarrow \nu_j^1=0$.}

In this case, we showed in Subsection 4.4 that 
$L_0(0)=T(0)={\tilde \nu}_j^1$ and $\lim_{\eta\rightarrow 0}{L_1}=
\kappa |\nu_j^2|+{\tilde \nu}_j^2$, 
where $\kappa=sign(\eta)$ and $\eta\rightarrow 0$ is 
a directional limit. Using the fact that $\eta L_2$ tends to 
$\kappa |\nu_j^2|$, equation (\ref{RpsiB}) gives:
\be
{\tilde g}^{(B)}_{\psi\psi}=\frac{4}{r^2f}
\frac{({\tilde \nu}_j^1)^2}{\sin^2 (\theta)} + O(\rho)~~,
\ee
so that the radius 
of the $\psi$-circle is finite for $\theta\neq 0, \pi$. Relations 
(\ref{tau_as}) imply:
\be
\label{las}
l=Re\tau=-\frac{{\tilde \nu}_j^2}{{\tilde \nu}_j^2}-
\kappa \frac{|\nu_j^2|}{{\tilde \nu}_j^1}(1-\frac{f}{2}\sin^2\theta)
+O(\rho)~~,~~
|Im\tau|=\infty~~.
\ee
Finally, equation (\ref{Bform}) gives:
\bea
\label{Blim}
B^{NS}_{\psi\rho}=\left[{\tilde \nu}_j^2+\kappa |\nu_j^2|\right]
\frac{\cos\chi}{\tan \theta}+O(\rho)~~&,&~~
B^{R}_{\psi\rho}=\left[\frac{(\nu_j^2)^2+
({\tilde \nu}_j^2)^2}{{\tilde \nu}_j^1}+
\kappa |\nu_j^2|\frac{2{\tilde \nu}_j^2}{{\tilde \nu}_j^1}\right]
\frac{\cos\chi}{\tan \theta}+O(\rho)~~\nn\\
B^{NS}_{\psi\chi}={\tilde \nu}_j^1+O(\rho)~~&,&~~
B^{R}_{\psi\chi}={\tilde \nu}_j^2+\kappa |\nu_j^2|+O(\rho)~~.
\eea

It may seem that the vertical locus supports a
weakly-coupled D7-brane whose worldvolume is spanned by the directions
of ${\bf E}^{3,1}$ and $r,\theta,\chi,\psi$. However recall from
(\ref{sqB}) that the $\chi$-circle is shrunk to a point 
for $\rho\rightarrow 0$. 
Hence we don't have enough dimensions for a
D7-brane, and we must interpret this 7- dimensional world-volume  
as a {\em delocalized D5-brane} (the delocalization is along the 
$\psi$-direction). 

To substantiate this interpretation, let us consider the same curve 
$\gamma$ as in Section 4.4. (figure \ref{curve}). Since the 
$\chi$-circle collapses for $\rho=0$, fibering it over $\gamma$ at a small 
but nonzero value of $\rho$ gives 
a surface $\Sigma$ similar to that of figure \ref{surface_a} (see 
figure \ref{surface_b} below). We have transverse coordinates  
$\rho,\eta$ and $\chi$, together with the delocalization direction
$\psi$ which we view as `transverse' as explained above. 
We shall 
compute the flux of the 3-form $F_3=d{\cal B}$ through a three-dimensional 
body $\Sigma'$ cutting the directions $\rho,\eta, \chi$ {\em and} $\psi$ and 
surrounding a point sitting at $\rho=\eta=0$ and at some 
fixed values of $r$ and $\theta$. 
For this, we choose the 
three-dimensional space $\Sigma'$ to be 
given by the $\psi$-circle fibration above $\Sigma$ (see figure 
\ref{surface_b}).

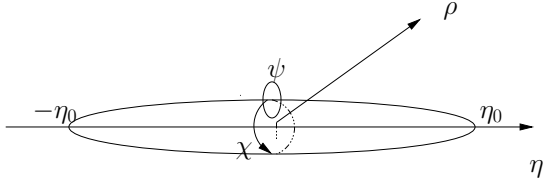
\begin{figure}[hbtp]
\begin{center}
\scalebox{0.6}{\input{surface_b.pstex_t}}
\end{center}
\caption{\label{surface_b} The 3-dimensional space $\Sigma'$ and the surface 
$\Sigma$ for type IIB vertical branes. The $\psi$-circle is fibered over the 
surface $\Sigma$ shown in the figure (this fibration gives $\Sigma'$). }
\end{figure}
\noindent

An argument similar to that of Section 4.4 shows that the flux of 
$F_3=d{\cal B}$ through $\Sigma'$ receives contributions only from the component $F_{\eta\psi\chi}$ (the contribution 
from the pull-back of $F_{\rho\chi\psi}$ vanishes since $\eta$ 
is very small and $d\rho=-2 \eta d\eta$ on $\Sigma'$). Equation (\ref{Bform}) 
shows that $F_{\eta\psi\chi}=\frac{\partial {\cal B}_\psi\chi}{\partial \eta}$
(to first order in $\rho$), so that the flux of interest is given by:
\be
\Phi=\left[\begin{array}{c}\Phi^{NS}\\\Phi^{R}\end{array}\right]=
\int_{\Sigma'}{d{\cal B}}=\int{d\chi\int{d\psi 
(\Delta {\cal B})_{\psi\chi}}}=\left[\begin{array}{c}0\\
2(2\pi)^2|\nu_j^2|\end{array}\right]~~.
\ee
To arrive at the last equation, we took the limit $\eta_0\rightarrow 0$
and used relations (\ref{Blim}). Notice that $(\Delta B^{NS})_{\psi\chi}=0$, 
which is why we obtain only a Ramond-Ramond flux:
\be
\label{Phi_Bv}
\Phi^{NS}=0~~,~~\Phi^{R}=2(2\pi)^2|\nu_j^2|~~.
\ee 

Once again, we obtain agreement with the interpretation proposed 
in \cite{toric}: when the isometry used for performing the IIA reduction 
fixes the vertical 
locus of interest in the twistor space, then the T-dual, 
IIB description contains a delocalized D5-brane along the associated type II 
locus. As discussed there, the gauge content predicted by (\ref{Phi_Bh}) and
(\ref{Phi_Bv})
must be `halved' in certain situations. As in Section 4, 
this can be understood as a consequence of possible 
identifications $\psi\rightarrow \psi+\pi$ above certain edges and 
vertices of $\Delta_M$.

\paragraph{Observation} Integrating the field strength $F_1=dl$ 
over the curve $\gamma$, one obtains:
\be
\int_{\gamma}{dl}=l(\eta_0)-l(-\eta_0)=\Delta l~~,
\ee
where $\Delta l=l(\eta=0^+)-l(\eta=0^-)$ is the discontinuity jump 
of $l$ at $\eta=0$. From equation (\ref{las}), we have:
\be
\Delta l=-\frac{2|\nu_j^2|}{{\tilde \nu}_j^1}
\left[1-\frac{f}{2}\sin^2(\theta)\right]~~.
\ee
However, this cannot be interpreted as the presence of  D7-brane charge. 
Indeed, such a charge must be computed by integration along a curve which 
surrounds a point on the D-brane worldvolume, which in our case 
can be obtained by considering two copies of $\gamma$ sitting at 
opposite angles $\chi$ and $\chi+\pi$ 
on the surface $\Sigma$ of figure \ref{surface_b}. 
Traversing the first curve from $-\eta_0$ to $\eta_0$ and the second from
$\eta_0$ to $-\eta_0$, one obtains a closed curve $\Gamma$ with the required 
property. The integral of $dl$ along $\Gamma$ equals $\Delta l -\Delta l=0$, 
so the object of interest does not carry  7-brane charge.

\section{The calibration $3$-form}
\setcounter{equation}{0}

We conclude with a calculation which is somewhat aside from the rest
of the paper, but may have independent applications, for example in
studies of $2$-dimensional $N=1$ superconformal 
field theories whose targets are spaces of
$G_2$ holonomy.\footnote{This is relevant for string propagation on
orbifolds of exceptional holonomy \cite{ShV}.} Namely we will
calculate the calibration three-form. 
Another reason for being interested in this object is the
issue of deformed $p$-brane solutions. In \cite{CLP}, it is shown that
one can obtain supersymmetric $p$-brane solutions supported by the
Chern-Simons term in the corresponding supergravity action if one
replaces the flat transverse space by a space admitting one or more
covariantly constant spinors\footnote{This guarantees that the
deformed brane is still supersymmetric.}. When the latter has $G_2$
holonomy, one obtains a deformed D$2$-brane solution in type IIA. Recall
that IIA supergravity contains a RR one-form and a RR three-form. 
If the former vanishes, then the D$2$-brane solution is completely
determined by a harmonic three form supported by the transverse space
\cite{CLP}.

In \cite{GP}, it is explained that the calibration $3$-form has the expression:
\be
\label{omega}
\omega = h_1 \omega_1 + h_2 \omega_2 + h_3 \omega_3 \, , \label{3form}
\ee
where $h_i$ depend only on $r$ and satisfy the following conditions:
\be
(h_1)^{\prime} = h_2 + h_3 \, , \qquad f^{-1/2} h_1 = \frac{1}{4} (r^2 f^{3/2} h_2)^{\prime} = (r^2 f^{-1/2} h_3)^{\prime} \, .
\ee
The $3$-forms $\omega_i$ are determined by the connection 
$A^j$ on the bundle $\Lambda^{2,-} (T^*M)$, whose curvature we denote by $F^j$:
\be
\omega_1 = \theta^i \wedge F^i\, , \qquad \omega_2 = dr \wedge (u^i F^i)\, , \qquad \omega_3 = dr \wedge \Sigma~~.
\ee
Here $\theta^i:= du^i + \epsilon^{ijk} A^j u^k$ and:
\be
\Sigma = \frac{1}{2} \epsilon_{ijk} u^i \theta^j \wedge \theta^k \, .
\ee
Using the relations (\ref{A}) for the connection one-form, we obtain:
\be \label{con}
A^1 = - \frac{F_{\eta}}{F} d\rho + \left(\frac{1}{2 \rho} +
\frac{F_{\rho}}{F} \right) d\eta~~,~~A^2 = - \frac{\sqrt{\rho}}{F}
d\phi~~,~~A^3 = \frac{\eta}{F\sqrt{\rho}} d\phi + \frac{1}{F
\sqrt{\rho}} d\psi \, . \ee
The curvature of this connection was calculated in Section 8 of \cite{CP}\footnote{Direct computation of the curvature of (\ref{con}) gives (\ref{curv}) upon using the fact that the function $F(\rho, \eta)$ is not arbitrary but satisfies $F_{\rho \rho} + F_{\eta \eta} = \frac{3 F}{4 \rho^2}$ \cite{CP}.}:
\bea \label{curv}
F^1 &=& - \, \frac{F^2 -4 \rho^2 (F_{\rho}^2 + F_{\eta}^2)}{4 F^2 \rho^2} \, 
d\rho \wedge d\eta + \frac{1}{F^2} \, d\phi \wedge d\psi \nn \\ 
F^2 &=& \frac{F_{\eta}}{F^2 \sqrt{\rho}} \, d\psi \wedge d\rho + 
\frac{1}{F^2 \sqrt{\rho}} \left(\rho F_{\rho} + \eta F_{\eta} -
\frac{F}{2} \right) d\phi \wedge d\rho \nn \\
&-& \frac{1}{F^2 \sqrt{\rho}} \left( F_{\rho} + \frac{F}{2 \rho} \right) \, d\psi \wedge d\eta + 
\frac{1}{F^2 \sqrt{\rho}} \left( \rho F_{\eta} - \eta F_{\rho} - \frac{\eta}{\rho} \frac{F}{2} \right) \, d\phi \wedge d\eta \nn
\\ F^3 &=& - \, \frac{1}{F^2 \sqrt{\rho}} \left(F_{\rho} +
\frac{F}{2 \rho}\right) d\psi \wedge d\rho + \frac{1}{F^2 \sqrt{\rho}} \left( \rho F_{\eta} - \eta F_{\rho} - \frac{\eta}{\rho} \frac{F}{2} \right) d\phi \wedge d\rho \nn
\\ &-& \frac{F_{\eta}}{F^2 \sqrt{\rho}} \, d\psi \wedge d\eta -
\frac{1}{F^2 \sqrt{\rho}} \left( \rho F_{\rho} + \eta F_{\eta} - \frac{F}{2} \right) d\phi
\wedge d\eta \, . \label{curvature} \eea It is now trivial to write
down $\omega_2$. To determine $\omega_3$, we calculate the quantity 
$\Sigma$: 
\bea
\Sigma &=& \left[\frac{F_{\eta}}{F} d\rho - \left(\frac{1}{2 \rho} +
\frac{F_{\rho}}{F}\right) d\eta \right] \wedge \left[\frac{(\rho u^3 + \eta
u^2)}{\sqrt{\rho} F} d\phi + \frac{u^2}{\sqrt{\rho} F} d\psi - du^1\right] -
\frac{u^1}{F^2} \, d\phi \wedge d\psi \nn \\ &+& \left[\frac{u^1
\sqrt{\rho}}{F} (u^1 du^2 - u^2 du^1) + \frac{u^1 \eta}{\sqrt{\rho} F}
(u^3 du^1 - u^1 du^3) + \frac{\rho u^3 + \eta u^2}{\sqrt{\rho} F} (u^3
du^2 - u^2 du^3)\right] \!\! \wedge \! d\phi \nn \\ &-& \frac{1}{F \sqrt{\rho}} \, du^3 \wedge d\psi + \frac{1}{2} \epsilon_{ijk} 
u^i du^j \wedge du^k \, .
\eea
Using (\ref{curvature}), we also compute:
{\footnotesize \bea
\omega_1 &=& \frac{1}{F^3 \sqrt{\rho}} \left\{ \left( \frac{F^2}{4 \rho^2} + 2 F_{\eta} F_{\rho} \right) (u^3 \rho + u^2 \eta) - (\rho + \eta) (u^2 + u^3) (F_{\rho}^2 + F_{\eta}^2) + \frac{F}{\rho} \left[ \eta (u^2 F_{\eta} - u^3 F_{\rho}) \right. \right. \nn \\
&-& \left. \left. \frac{F}{\rho} (u^3 \eta - u^2 \rho) \right] \right\} d\phi \wedge d\rho \wedge d\eta \, + \, \frac{u^1}{\rho F^3} \left( 2 \rho F_{\rho} + \eta F_{\eta} \right) d\psi \wedge d\phi \wedge d\rho \nn \\
&+& \frac{1}{\sqrt{\rho} F^3} \left\{ u^3 \left[ F_{\eta}^2 - \left( F_{\rho} + \frac{F}{2 \rho} \right)^2 \right] + u^2 \left( \frac{F^2}{4 \rho^2} - (F_{\rho} - F_{\eta})^2 + \frac{F_{\eta} F}{\rho} \right) \right\} d\psi \wedge d\rho \wedge d\eta \nn \\
&+& \frac{u^1}{\rho F^3} \left( 2 \rho F_{\eta} - \eta F_{\rho} - \frac{\eta}{\rho} \frac{F}{2} \right) d\psi \wedge d\phi \wedge d\eta + du^i \wedge F^i \, .
\eea}
The calibration three-form is now obtained by substituting these expressions 
into (\ref{omega}).

\section{Conclusions}

Upon using the recent result of \cite{CP}, we constructed explicit
metrics for a discrete infinity of $G_2$ cones (and their one-parameter 
deformations) associated with toric hyperkahler cones. 
This allowed us to perform the type 
IIA reduction of the general such model, 
and extract its IIB dual. This leads to a large class of solutions 
of IIA and IIB supergravity which preserve $N=1$ supersymmetry in the 
four external directions. By studying the asymptotics of the various fields 
in the vicinity of certain locations, we extracted the physical 
interpretation of such backgrounds. In particular, we showed that 
the resulting IIA 
solutions correspond to systems of weakly and strongly coupled D6-branes, 
while their type IIB duals describe systems of localized and delocalized 
5-branes. For particular values of the discrete parameters 
characterizing our metrics (namely the toric hyperkahler generators 
$\nu_j$ of \cite{toric}, which we reviewed in Section 2), one can perform 
reduction through a `good' isometry. In this case, we showed that
the type IIA solutions reduce to backgrounds 
containing only weakly coupled D6-branes, whose 
type IIB duals are delocalized, 
weakly coupled D5-branes. This provides 
an independent confirmation of the conclusions of \cite{toric}.

These T-dual IIA and IIB solutions form a wide generalization of systems 
studied in the recent work of \cite{Witten_Acharya}, and provide a rich 
source of $N=1$ type II vacua 
which admit an explicitly known lift to M-theory 
backgrounds of $G_2$ holonomy. The choices of $\nu_j$ which lead 
to type IIA solutions for weakly-coupled D6-branes  are of 
particular interest for phenomenological applications.

\acknowledgments{ The authors thank M.~Rocek for interest and encouragement.
C. I. L would like to thank the CERN Theory group  
for hospitality during the last stages in the preparation of this paper.
The present work was supported by the Research Foundation under NSF 
grant PHY-0098527.}

\end{document}

%% file: DDM.pstex_t
\begin{picture}(0,0)%
\epsfig{file=DDM.pstex}%
\end{picture}%
\setlength{\unitlength}{4144sp}%
\begingroup\makeatletter\ifx\SetFigFont\undefined%
\gdef\SetFigFont#1#2#3#4#5{%
  \reset@font\fontsize{#1}{#2pt}%
  \fontfamily{#3}\fontseries{#4}\fontshape{#5}%
  \selectfont}%
\fi\endgroup%
\begin{picture}(8713,4063)(2010,-4381)
\put(9316,-4246){\makebox(0,0)[lb]{\smash{\SetFigFont{17}{20.4}{\rmdefault}{\bfdefault}{\updefault}\special{ps: gsave 0 0 0 setrgbcolor}$\Delta_M$\special{ps: grestore}}}}
\put(4006,-4381){\makebox(0,0)[lb]{\smash{\SetFigFont{17}{20.4}{\rmdefault}{\bfdefault}{\updefault}\special{ps: gsave 0 0 0 setrgbcolor}$\Delta$\special{ps: grestore}}}}
\put(3421,-781){\makebox(0,0)[lb]{\smash{\SetFigFont{17}{20.4}{\rmdefault}{\bfdefault}{\updefault}\special{ps: gsave 0 0 0 setrgbcolor}$D_1$\special{ps: grestore}}}}
\put(5851,-1051){\makebox(0,0)[lb]{\smash{\SetFigFont{17}{20.4}{\rmdefault}{\bfdefault}{\updefault}\special{ps: gsave 0 0 0 setrgbcolor}$p_e$\special{ps: grestore}}}}
\put(5086,-2851){\makebox(0,0)[lb]{\smash{\SetFigFont{17}{20.4}{\rmdefault}{\bfdefault}{\updefault}\special{ps: gsave 0 0 0 setrgbcolor}$D_3$\special{ps: grestore}}}}
\put(5131,-1906){\makebox(0,0)[lb]{\smash{\SetFigFont{17}{20.4}{\rmdefault}{\bfdefault}{\updefault}\special{ps: gsave 0 0 0 setrgbcolor}$D_2$\special{ps: grestore}}}}
\put(4321,-1231){\makebox(0,0)[lb]{\smash{\SetFigFont{17}{20.4}{\rmdefault}{\bfdefault}{\updefault}\special{ps: gsave 0 0 0 setrgbcolor}$D_4$\special{ps: grestore}}}}
\end{picture}

%% file: dist.pstex_t
\begin{picture}(0,0)%
\epsfig{file=dist.pstex}%
\end{picture}%
\setlength{\unitlength}{4144sp}%
\begingroup\makeatletter\ifx\SetFigFont\undefined%
\gdef\SetFigFont#1#2#3#4#5{%
  \reset@font\fontsize{#1}{#2pt}%
  \fontfamily{#3}\fontseries{#4}\fontshape{#5}%
  \selectfont}%
\fi\endgroup%
\begin{picture}(4600,4237)(1936,-4381)
\put(4006,-4381){\makebox(0,0)[lb]{\smash{\SetFigFont{17}{20.4}{\rmdefault}{\bfdefault}{\updefault}\special{ps: gsave 0 0 0 setrgbcolor}$\Delta$\special{ps: grestore}}}}
\end{picture}

%% file: CO.pstex_t
\begin{picture}(0,0)%
\includegraphics{CO.pstex}%
\end{picture}%
\setlength{\unitlength}{4144sp}%
\begingroup\makeatletter\ifx\SetFigFont\undefined%
\gdef\SetFigFont#1#2#3#4#5{%
  \reset@font\fontsize{#1}{#2pt}%
  \fontfamily{#3}\fontseries{#4}\fontshape{#5}%
  \selectfont}%
\fi\endgroup%
\begin{picture}(4095,2625)(946,-2221)
\put(4636,-1906){\makebox(0,0)[lb]{\smash{\SetFigFont{17}{20.4}{\rmdefault}{\bfdefault}{\updefault}\special{ps: gsave 0 0 0 setrgbcolor}${\overline {\cal H}}^2$\special{ps: grestore}}}}
\put(1171,-16){\makebox(0,0)[lb]{\smash{\SetFigFont{17}{20.4}{\rmdefault}{\bfdefault}{\updefault}\special{ps: gsave 0 0 0 setrgbcolor}$X$\special{ps: grestore}}}}
\put(2836,164){\makebox(0,0)[lb]{\smash{\SetFigFont{17}{20.4}{\rmdefault}{\bfdefault}{\updefault}\special{ps: gsave 0 0 0 setrgbcolor}$\H^*$\special{ps: grestore}}}}
\put(5041,-871){\makebox(0,0)[lb]{\smash{\SetFigFont{17}{20.4}{\rmdefault}{\bfdefault}{\updefault}\special{ps: gsave 0 0 0 setrgbcolor}$\pi_M$\special{ps: grestore}}}}
\put(2746,-2221){\makebox(0,0)[lb]{\smash{\SetFigFont{17}{20.4}{\rmdefault}{\bfdefault}{\updefault}\special{ps: gsave 0 0 0 setrgbcolor}$\Theta$\special{ps: grestore}}}}
\put(1171,-1951){\makebox(0,0)[lb]{\smash{\SetFigFont{17}{20.4}{\rmdefault}{\bfdefault}{\updefault}\special{ps: gsave 0 0 0 setrgbcolor}$\R^6$\special{ps: grestore}}}}
\put(4681,-16){\makebox(0,0)[lb]{\smash{\SetFigFont{17}{20.4}{\rmdefault}{\bfdefault}{\updefault}\special{ps: gsave 0 0 0 setrgbcolor}$M$\special{ps: grestore}}}}
\put(946,-916){\makebox(0,0)[lb]{\smash{\SetFigFont{17}{20.4}{\rmdefault}{\bfdefault}{\updefault}\special{ps: gsave 0 0 0 setrgbcolor}${\vec \pi}$\special{ps: grestore}}}}
\end{picture}

%% file: sph.pstex_t
\begin{picture}(0,0)%
\epsfig{file=sph.pstex}%
\end{picture}%
\setlength{\unitlength}{4144sp}%
\begingroup\makeatletter\ifx\SetFigFont\undefined%
\gdef\SetFigFont#1#2#3#4#5{%
  \reset@font\fontsize{#1}{#2pt}%
  \fontfamily{#3}\fontseries{#4}\fontshape{#5}%
  \selectfont}%
\fi\endgroup%
\begin{picture}(2970,2916)(1081,-3112)
\put(1756,-466){\makebox(0,0)[lb]{\smash{\SetFigFont{17}{20.4}{\rmdefault}{\bfdefault}{\updefault}\special{ps: gsave 0 0 0 setrgbcolor}$P_1$\special{ps: grestore}}}}
\put(2026,-1636){\makebox(0,0)[lb]{\smash{\SetFigFont{17}{20.4}{\rmdefault}{\bfdefault}{\updefault}\special{ps: gsave 0 0 0 setrgbcolor}$\Delta_M$\special{ps: grestore}}}}
\put(2026,-3031){\makebox(0,0)[lb]{\smash{\SetFigFont{17}{20.4}{\rmdefault}{\bfdefault}{\updefault}\special{ps: gsave 0 0 0 setrgbcolor}$\partial {\overline {\cal H}}^2$\special{ps: grestore}}}}
\put(1081,-2266){\makebox(0,0)[lb]{\smash{\SetFigFont{17}{20.4}{\rmdefault}{\bfdefault}{\updefault}\special{ps: gsave 0 0 0 setrgbcolor}$P_3$\special{ps: grestore}}}}
\put(3826,-2356){\makebox(0,0)[lb]{\smash{\SetFigFont{17}{20.4}{\rmdefault}{\bfdefault}{\updefault}\special{ps: gsave 0 0 0 setrgbcolor}$P_2$\special{ps: grestore}}}}
\put(4051,-1276){\makebox(0,0)[lb]{\smash{\SetFigFont{17}{20.4}{\rmdefault}{\bfdefault}{\updefault}\special{ps: gsave 0 0 0 setrgbcolor}$P_4$\special{ps: grestore}}}}
\end{picture}

%% file: ctorus.pstex_t
\begin{picture}(0,0)%
\epsfig{file=ctorus.pstex}%
\end{picture}%
\setlength{\unitlength}{4144sp}%
\begingroup\makeatletter\ifx\SetFigFont\undefined%
\gdef\SetFigFont#1#2#3#4#5{%
  \reset@font\fontsize{#1}{#2pt}%
  \fontfamily{#3}\fontseries{#4}\fontshape{#5}%
  \selectfont}%
\fi\endgroup%
\begin{picture}(6402,5001)(2161,-5452)
\put(2746,-3481){\makebox(0,0)[lb]{\smash{\SetFigFont{17}{20.4}{\rmdefault}{\bfdefault}{\updefault}\special{ps: gsave 0 0 0 setrgbcolor}$\zeta$\special{ps: grestore}}}}
\put(7516,-5191){\makebox(0,0)[lb]{\smash{\SetFigFont{17}{20.4}{\rmdefault}{\bfdefault}{\updefault}\special{ps: gsave 0 0 0 setrgbcolor}$\xi$\special{ps: grestore}}}}
\put(2746,-691){\makebox(0,0)[lb]{\smash{\SetFigFont{17}{20.4}{\rmdefault}{\bfdefault}{\updefault}\special{ps: gsave 0 0 0 setrgbcolor}$\zeta$\special{ps: grestore}}}}
\put(7516,-2401){\makebox(0,0)[lb]{\smash{\SetFigFont{17}{20.4}{\rmdefault}{\bfdefault}{\updefault}\special{ps: gsave 0 0 0 setrgbcolor}$\xi$\special{ps: grestore}}}}
\put(4141,-3031){\makebox(0,0)[lb]{\smash{\SetFigFont{17}{20.4}{\rmdefault}{\bfdefault}{\updefault}\special{ps: gsave 0 0 0 setrgbcolor}$2\pi$\special{ps: grestore}}}}
\put(6751,-3166){\makebox(0,0)[lb]{\smash{\SetFigFont{17}{20.4}{\rmdefault}{\bfdefault}{\updefault}\special{ps: gsave 0 0 0 setrgbcolor}$2\pi T$\special{ps: grestore}}}}
\put(2161,-961){\makebox(0,0)[lb]{\smash{\SetFigFont{17}{20.4}{\rmdefault}{\bfdefault}{\updefault}\special{ps: gsave 0 0 0 setrgbcolor}$2\pi$\special{ps: grestore}}}}
\put(3196,-2581){\makebox(0,0)[lb]{\smash{\SetFigFont{17}{20.4}{\rmdefault}{\bfdefault}{\updefault}\special{ps: gsave 0 0 0 setrgbcolor}$\Gamma_\xi=\Gamma_\psi$\special{ps: grestore}}}}
\put(2341,-1636){\makebox(0,0)[lb]{\smash{\SetFigFont{17}{20.4}{\rmdefault}{\bfdefault}{\updefault}\special{ps: gsave 0 0 0 setrgbcolor}$\Gamma_\zeta$\special{ps: grestore}}}}
\put(4411,-1366){\makebox(0,0)[lb]{\smash{\SetFigFont{17}{20.4}{\rmdefault}{\bfdefault}{\updefault}\special{ps: gsave 0 0 0 setrgbcolor}$\Gamma_\chi$\special{ps: grestore}}}}
\put(3196,-5371){\makebox(0,0)[lb]{\smash{\SetFigFont{17}{20.4}{\rmdefault}{\bfdefault}{\updefault}\special{ps: gsave 0 0 0 setrgbcolor}$\Gamma_\xi=\Gamma_\psi$\special{ps: grestore}}}}
\put(4816,-4876){\makebox(0,0)[lb]{\smash{\SetFigFont{17}{20.4}{\rmdefault}{\bfdefault}{\updefault}\special{ps: gsave 0 0 0 setrgbcolor}$\Gamma_\chi$\special{ps: grestore}}}}
\put(2386,-4921){\makebox(0,0)[lb]{\smash{\SetFigFont{17}{20.4}{\rmdefault}{\bfdefault}{\updefault}\special{ps: gsave 0 0 0 setrgbcolor}$\Gamma_\zeta$\special{ps: grestore}}}}
\end{picture}

%% file: cycle.pstex_t
\begin{picture}(0,0)%
\epsfig{file=cycle.pstex}%
\end{picture}%
\setlength{\unitlength}{4144sp}%
\begingroup\makeatletter\ifx\SetFigFont\undefined%
\gdef\SetFigFont#1#2#3#4#5{%
  \reset@font\fontsize{#1}{#2pt}%
  \fontfamily{#3}\fontseries{#4}\fontshape{#5}%
  \selectfont}%
\fi\endgroup%
\begin{picture}(5865,5760)(766,-5236)
\put(766, 29){\makebox(0,0)[lb]{\smash{\SetFigFont{17}{20.4}{\rmdefault}{\bfdefault}{\updefault}\special{ps: gsave 0 0 0 setrgbcolor}$\Gamma_\xi$\special{ps: grestore}}}}
\put(4546,254){\makebox(0,0)[lb]{\smash{\SetFigFont{17}{20.4}{\rmdefault}{\bfdefault}{\updefault}\special{ps: gsave 0 0 0 setrgbcolor}$\Gamma_\zeta$\special{ps: grestore}}}}
\put(991,-5236){\makebox(0,0)[lb]{\smash{\SetFigFont{17}{20.4}{\rmdefault}{\bfdefault}{\updefault}\special{ps: gsave 0 0 0 setrgbcolor}$\Gamma_\chi$\special{ps: grestore}}}}
\end{picture}

%% file: hor_strong_a.pstex_t
\begin{picture}(0,0)%
\epsfig{file=hor_strong_a.pstex}%
\end{picture}%
\setlength{\unitlength}{4144sp}%
\begingroup\makeatletter\ifx\SetFigFont\undefined%
\gdef\SetFigFont#1#2#3#4#5{%
  \reset@font\fontsize{#1}{#2pt}%
  \fontfamily{#3}\fontseries{#4}\fontshape{#5}%
  \selectfont}%
\fi\endgroup%
\begin{picture}(8520,5209)(181,-6557)
\put(5166,-5356){\makebox(0,0)[lb]{\smash{\SetFigFont{14}{16.8}{\rmdefault}{\bfdefault}{\updefault}\special{ps: gsave 0 0 0 setrgbcolor}$\theta$\special{ps: grestore}}}}
\put(4123,-5207){\makebox(0,0)[lb]{\smash{\SetFigFont{14}{16.8}{\rmdefault}{\bfdefault}{\updefault}\special{ps: gsave 0 0 0 setrgbcolor}$\xi$\special{ps: grestore}}}}
\put(3736,-2851){\makebox(0,0)[lb]{\smash{\SetFigFont{20}{24.0}{\rmdefault}{\bfdefault}{\updefault}\special{ps: gsave 0 0 0 setrgbcolor}$\theta=\pi$\special{ps: grestore}}}}
\put(3736,-1591){\makebox(0,0)[lb]{\smash{\SetFigFont{20}{24.0}{\rmdefault}{\bfdefault}{\updefault}\special{ps: gsave 0 0 0 setrgbcolor}$\theta=0$\special{ps: grestore}}}}
\put(1216,-3436){\makebox(0,0)[lb]{\smash{\SetFigFont{17}{20.4}{\rmdefault}{\bfdefault}{\updefault}\special{ps: gsave 0 0 0 setrgbcolor}$\eta$\special{ps: grestore}}}}
\put(181,-3256){\makebox(0,0)[lb]{\smash{\SetFigFont{17}{20.4}{\rmdefault}{\bfdefault}{\updefault}\special{ps: gsave 0 0 0 setrgbcolor}$\eta_i\neq 0$\special{ps: grestore}}}}
\put(5671,-3391){\makebox(0,0)[lb]{\smash{\SetFigFont{17}{20.4}{\rmdefault}{\bfdefault}{\updefault}\special{ps: gsave 0 0 0 setrgbcolor}$\eta_i\neq 0$\special{ps: grestore}}}}
\put(6886,-3526){\makebox(0,0)[lb]{\smash{\SetFigFont{17}{20.4}{\rmdefault}{\bfdefault}{\updefault}\special{ps: gsave 0 0 0 setrgbcolor}$\eta$\special{ps: grestore}}}}
\put(2746,-3301){\makebox(0,0)[lb]{\smash{\SetFigFont{17}{20.4}{\rmdefault}{\bfdefault}{\updefault}\special{ps: gsave 0 0 0 setrgbcolor}$\eta_j\neq 0$\special{ps: grestore}}}}
\put(8371,-3346){\makebox(0,0)[lb]{\smash{\SetFigFont{17}{20.4}{\rmdefault}{\bfdefault}{\updefault}\special{ps: gsave 0 0 0 setrgbcolor}$\eta_j=0$\special{ps: grestore}}}}
\put(4231,-2221){\makebox(0,0)[lb]{\smash{\SetFigFont{17}{20.4}{\rmdefault}{\bfdefault}{\updefault}\special{ps: gsave 0 0 0 setrgbcolor}$\theta$\special{ps: grestore}}}}
\end{picture}

%% file: hlocus.pstex_t
\begin{picture}(0,0)%
\epsfig{file=hlocus.pstex}%
\end{picture}%
\setlength{\unitlength}{4144sp}%
\begingroup\makeatletter\ifx\SetFigFont\undefined%
\gdef\SetFigFont#1#2#3#4#5{%
  \reset@font\fontsize{#1}{#2pt}%
  \fontfamily{#3}\fontseries{#4}\fontshape{#5}%
  \selectfont}%
\fi\endgroup%
\begin{picture}(4666,4140)(1050,-4066)
\put(5716,-1726){\makebox(0,0)[lb]{\smash{\SetFigFont{14}{16.8}{\rmdefault}{\bfdefault}{\updefault}\special{ps: gsave 0 0 0 setrgbcolor}$\theta$\special{ps: grestore}}}}
\put(2656,-4066){\makebox(0,0)[lb]{\smash{\SetFigFont{14}{16.8}{\rmdefault}{\bfdefault}{\updefault}\special{ps: gsave 0 0 0 setrgbcolor}$\eta$\special{ps: grestore}}}}
\put(2431,-106){\makebox(0,0)[lb]{\smash{\SetFigFont{14}{16.8}{\rmdefault}{\bfdefault}{\updefault}\special{ps: gsave 0 0 0 setrgbcolor}$\xi$\special{ps: grestore}}}}
\end{picture}

%% file: curve.pstex_t
\begin{picture}(0,0)%
\epsfig{file=curve.pstex}%
\end{picture}%
\setlength{\unitlength}{4144sp}%
\begingroup\makeatletter\ifx\SetFigFont\undefined%
\gdef\SetFigFont#1#2#3#4#5{%
  \reset@font\fontsize{#1}{#2pt}%
  \fontfamily{#3}\fontseries{#4}\fontshape{#5}%
  \selectfont}%
\fi\endgroup%
\begin{picture}(5784,7275)(1969,-6826)
\put(7291,-3526){\makebox(0,0)[lb]{\smash{\SetFigFont{17}{20.4}{\rmdefault}{\bfdefault}{\updefault}\special{ps: gsave 0 0 0 setrgbcolor}$\eta$\special{ps: grestore}}}}
\put(6751,-3166){\makebox(0,0)[lb]{\smash{\SetFigFont{17}{20.4}{\rmdefault}{\bfdefault}{\updefault}\special{ps: gsave 0 0 0 setrgbcolor}$\eta_0$\special{ps: grestore}}}}
\put(2746,-3256){\makebox(0,0)[lb]{\smash{\SetFigFont{17}{20.4}{\rmdefault}{\bfdefault}{\updefault}\special{ps: gsave 0 0 0 setrgbcolor}$-\eta_0$\special{ps: grestore}}}}
\put(6346,-1771){\makebox(0,0)[lb]{\smash{\SetFigFont{17}{20.4}{\rmdefault}{\bfdefault}{\updefault}\special{ps: gsave 0 0 0 setrgbcolor}$\gamma$\special{ps: grestore}}}}
\put(4681,209){\makebox(0,0)[lb]{\smash{\SetFigFont{17}{20.4}{\rmdefault}{\bfdefault}{\updefault}\special{ps: gsave 0 0 0 setrgbcolor}$\rho$\special{ps: grestore}}}}
\put(5041,-6406){\makebox(0,0)[lb]{\smash{\SetFigFont{17}{20.4}{\rmdefault}{\bfdefault}{\updefault}\special{ps: gsave 0 0 0 setrgbcolor}$\gamma$\special{ps: grestore}}}}
\put(5176,-3706){\makebox(0,0)[lb]{\smash{\SetFigFont{17}{20.4}{\rmdefault}{\bfdefault}{\updefault}\special{ps: gsave 0 0 0 setrgbcolor}$(\rho\rightarrow 0)$\special{ps: grestore}}}}
\put(4951,-1366){\makebox(0,0)[lb]{\smash{\SetFigFont{17}{20.4}{\rmdefault}{\bfdefault}{\updefault}\special{ps: gsave 0 0 0 setrgbcolor}$\eta_0^2$\special{ps: grestore}}}}
\end{picture}

%% file: surface_a.pstex_t
\begin{picture}(0,0)%
\epsfig{file=surface_a.pstex}%
\end{picture}%
\setlength{\unitlength}{4144sp}%
\begingroup\makeatletter\ifx\SetFigFont\undefined%
\gdef\SetFigFont#1#2#3#4#5{%
  \reset@font\fontsize{#1}{#2pt}%
  \fontfamily{#3}\fontseries{#4}\fontshape{#5}%
  \selectfont}%
\fi\endgroup%
\begin{picture}(5289,1632)(2104,-4111)
\put(7336,-4111){\makebox(0,0)[lb]{\smash{\SetFigFont{17}{20.4}{\rmdefault}{\bfdefault}{\updefault}\special{ps: gsave 0 0 0 setrgbcolor}$\eta$\special{ps: grestore}}}}
\put(6841,-3571){\makebox(0,0)[lb]{\smash{\SetFigFont{17}{20.4}{\rmdefault}{\bfdefault}{\updefault}\special{ps: gsave 0 0 0 setrgbcolor}$\eta_0$\special{ps: grestore}}}}
\put(2386,-3571){\makebox(0,0)[lb]{\smash{\SetFigFont{17}{20.4}{\rmdefault}{\bfdefault}{\updefault}\special{ps: gsave 0 0 0 setrgbcolor}$-\eta_0$\special{ps: grestore}}}}
\put(4141,-3931){\makebox(0,0)[lb]{\smash{\SetFigFont{17}{20.4}{\rmdefault}{\bfdefault}{\updefault}\special{ps: gsave 0 0 0 setrgbcolor}$\zeta$\special{ps: grestore}}}}
\put(4501,-2941){\makebox(0,0)[lb]{\smash{\SetFigFont{17}{20.4}{\rmdefault}{\bfdefault}{\updefault}\special{ps: gsave 0 0 0 setrgbcolor}$\rho$\special{ps: grestore}}}}
\end{picture}

%% file: b_degeneration.pstex_t
\begin{picture}(0,0)%
\includegraphics{b_degeneration.pstex}%
\end{picture}%
\setlength{\unitlength}{4144sp}%
\begingroup\makeatletter\ifx\SetFigFont\undefined%
\gdef\SetFigFont#1#2#3#4#5{%
  \reset@font\fontsize{#1}{#2pt}%
  \fontfamily{#3}\fontseries{#4}\fontshape{#5}%
  \selectfont}%
\fi\endgroup%
\begin{picture}(10497,6069)(361,-6208)
\put(3376,-6001){\makebox(0,0)[lb]{\smash{\SetFigFont{17}{20.4}{\rmdefault}{\bfdefault}{\updefault}{$\eta_i$}%
}}}
\put(8281,-6046){\makebox(0,0)[lb]{\smash{\SetFigFont{17}{20.4}{\rmdefault}{\bfdefault}{\updefault}{$\eta_j$}%
}}}
\put(4771,-1276){\makebox(0,0)[lb]{\smash{\SetFigFont{17}{20.4}{\rmdefault}{\bfdefault}{\updefault}{$\eta$}%
}}}
\put(361,-1951){\makebox(0,0)[lb]{\smash{\SetFigFont{17}{20.4}{\rmdefault}{\bfdefault}{\updefault}{$\eta_i\neq 0$}%
}}}
\put(4231,-1951){\makebox(0,0)[lb]{\smash{\SetFigFont{17}{20.4}{\rmdefault}{\bfdefault}{\updefault}{$\eta_j\neq 0$}%
}}}
\put(5851,-331){\makebox(0,0)[lb]{\smash{\SetFigFont{17}{20.4}{\rmdefault}{\bfdefault}{\updefault}{$\theta$}%
}}}
\put(8506,-2716){\makebox(0,0)[lb]{\smash{\SetFigFont{17}{20.4}{\rmdefault}{\bfdefault}{\updefault}{$T\neq 0$}%
}}}
\put(2251,-2761){\makebox(0,0)[lb]{\smash{\SetFigFont{17}{20.4}{\rmdefault}{\bfdefault}{\updefault}{$T\neq 0$}%
}}}
\put(5851,-6136){\makebox(0,0)[lb]{\smash{\SetFigFont{17}{20.4}{\rmdefault}{\bfdefault}{\updefault}{$T= 0$}%
}}}
\put(6616,-1996){\makebox(0,0)[lb]{\smash{\SetFigFont{17}{20.4}{\rmdefault}{\bfdefault}{\updefault}{$\eta_i= 0$}%
}}}
\put(7246,-3481){\makebox(0,0)[lb]{\smash{\SetFigFont{17}{20.4}{\rmdefault}{\bfdefault}{\updefault}{$\psi$}%
}}}
\put(3376,-2536){\makebox(0,0)[lb]{\smash{\SetFigFont{17}{20.4}{\rmdefault}{\bfdefault}{\updefault}{$\psi$}%
}}}
\put(10486,-1996){\makebox(0,0)[lb]{\smash{\SetFigFont{17}{20.4}{\rmdefault}{\bfdefault}{\updefault}{$\eta_j\neq 0$}%
}}}
\end{picture}

%% file: surface_b.pstex_t
\begin{picture}(0,0)%
\includegraphics{surface_b.pstex}%
\end{picture}%
\setlength{\unitlength}{4144sp}%
\begingroup\makeatletter\ifx\SetFigFont\undefined%
\gdef\SetFigFont#1#2#3#4#5{%
  \reset@font\fontsize{#1}{#2pt}%
  \fontfamily{#3}\fontseries{#4}\fontshape{#5}%
  \selectfont}%
\fi\endgroup%
\begin{picture}(5289,1839)(2104,-4183)
\put(7336,-4111){\makebox(0,0)[lb]{\smash{\SetFigFont{17}{20.4}{\rmdefault}{\bfdefault}{\updefault}{$\eta$}%
}}}
\put(6841,-3571){\makebox(0,0)[lb]{\smash{\SetFigFont{17}{20.4}{\rmdefault}{\bfdefault}{\updefault}{$\eta_0$}%
}}}
\put(2386,-3571){\makebox(0,0)[lb]{\smash{\SetFigFont{17}{20.4}{\rmdefault}{\bfdefault}{\updefault}{$-\eta_0$}%
}}}
\put(4726,-3166){\makebox(0,0)[lb]{\smash{\SetFigFont{17}{20.4}{\rmdefault}{\bfdefault}{\updefault}{$\psi$}%
}}}
\put(4411,-3931){\makebox(0,0)[lb]{\smash{\SetFigFont{17}{20.4}{\rmdefault}{\bfdefault}{\updefault}{$\chi$}%
}}}
\put(6481,-2536){\makebox(0,0)[lb]{\smash{\SetFigFont{17}{20.4}{\rmdefault}{\bfdefault}{\updefault}{$\rho$}%
}}}
\end{picture}